%% file: Rigault_SF_bias_H0_paper.tex
\documentclass[10pt,aasms,preprint]{aastex}

\usepackage{amsmath}
\usepackage{lineno}
\usepackage{ulem}

\newcommand{\Hnot}{\ensuremath{H_0}}
\newcommand{\Hnotcorr}{\ensuremath{H_0^{\mathrm{corr}}}}

\newcommand{\SFRSD}{\ensuremath{\Sigma_{\mathrm{SFR}}}}
\newcommand{\sSFR}{\ensuremath{\mathrm{sSFR}}}
\newcommand{\pIe}{\ensuremath{\mathcal{P}\mathrm{(Ia\epsilon)}}}
\newcommand{\pIa}{\ensuremath{\mathrm{\mathcal{P}({Ia}\alpha)}}}
\newcommand{\shoes}{SH0ES11}
\newcommand{\constitution}{\textrm{Constitution}}
\newcommand{\SNfactory}{\textrm{SNfactory}}
\newcommand{\GALEX}{\textit{GALEX}}

\input{Number_H0paper}

\newcommand{\Ha}{\ensuremath{\mathrm{H}\alpha}}
\newcommand{\SigmaHa}{\ensuremath{\Sigma_{\Ha}}}
\newcommand{\sS}{\ensuremath{\mathrm{Ia}\alpha}}
\newcommand{\sN}{\ensuremath{\mathrm{Ia}\epsilon}}
\newcommand{\cHR}{\ensuremath{\Delta M_B^{\mathrm{corr}}}}
\newcommand{\DcHR}{%
  \ensuremath{\delta \langle M_B^{\mathrm{corr}}\rangle_{\mathrm{SF}}}}

\shorttitle{SN~Ia Environment Bias Affecting \Hnot{}} 
\shortauthors{M.~Rigault and the Nearby Supernova Factory}

\begin{document}

\title{Confirmation of a Star Formation Bias in
 Type Ia Supernova \\Distances 
 and its Effect on Measurement of the Hubble Constant}
\slugcomment{{\sc Submitted to ApJ}: Oct.~30, 2014 -- {\sc Accepted}: Dec.~17, 2014}

\input{authors}
\begin{abstract}
Previously we used the Nearby Supernova Factory sample
to show that SNe~Ia having locally star-forming environments are dimmer than
SNe~Ia having locally passive environments.
Here we use the \constitution\ sample together with host galaxy data
from \GALEX\ to independently confirm that result.
The effect is seen
using both the SALT2 and MLCS2k2 lightcurve fitting and standardization
methods, with brightness
differences of \SFHickenSALT{} for SALT2 and \SFHickenMLCSRiess{} 
for MLCS2k2 with $R_V=2.5$.
When combined with our previous measurement the effect
is \SFcombinedSALT{} for SALT2. If the ratio of these local SN~Ia
environments changes with redshift or sample 
selection, this can lead to a bias in cosmological measurements. 
We explore this issue further, using as an example
the direct measurement of \Hnot.
\GALEX{} observations show that the SNe~Ia having standardized
absolute magnitudes calibrated via the Cepheid period--luminosity
relation using {\textit{HST}}
originate in predominately star-forming
environments, whereas only $\sim$\,50\% of the Hubble-flow comparison
sample have locally star-forming environments. As a consequence,
the \Hnot{} measurement using SNe~Ia is currently overestimated. 
Correcting for this bias, we find a value of
$\Hnotcorr=$~\HnotCorrEftaTMLCSRiessMassRmv{} when using the LMC distance,
Milky Way parallaxes and the NGC~4258 megamaser as the Cepheid
zeropoint, and
\HnotCorrEftaOMLCSRiessMassRmv{} when only using NGC~4258.
Our correction brings the direct measurement
of \Hnot{} within $\sim 1\,\sigma$ of recent indirect measurements
based on the CMB power spectrum.

\end{abstract}


\section{Introduction}

Empirically-standardized Type Ia supernovae (SNe~Ia) have been
developed into powerful distance indicators. Their use in deriving
the expansion history of the Universe led to the discovery of the
acceleration of the cosmological expansion
\citep{perlmutter_measurements_1999, riess_observational_1998}.
They also have proven to be important in accurately measuring the
local Hubble constant, \Hnot. The sample of events in nearby host
galaxies within the range of Cepheid and maser calibration is
growing, and these can be coupled to other SNe~Ia at redshifts where
host peculiar motions are negligible in comparison to the current
cosmic expansion rate.  For instance the SH0ES program \citep[][hereafter
\shoes]{riess_3_2011} has reached a quoted precision of 3\% on the
measurement of \Hnot{} using SNe~Ia, reporting a value of $73.8
\pm~2.4\ \mathrm{km\ s^{-1}\ Mpc^{-1}}$.  Subsequently,
\citet{humphreys_correction_2013} adjusted the distance to the
NGC~4258 megamaser, used as one of the Cepheid zero-points by \shoes,
which reduced \Hnot\ to $72.7 \pm2.4\ \mathrm{km\ s^{-1}\ Mpc^{-1}}$.
More recently, \cite{Efstathiou_H0_2014} re-examined the Cepheid
analysis of \shoes{} and made an additional small adjustment, to
$\Hnot=72.5 \pm2.5\ \mathrm{km\ s^{-1}\ Mpc^{-1}}$.
Similarly, the HST Key Project and Carnegie Hubble Project
\citep{freedman_final_2001, freedman_2012} have relied heavily on
SNe~Ia to obtain their result of $74.3\pm2.1\ \mathrm{km\ s^{-1}\ Mpc^{-1}}$.

The value of \Hnot{} has become the center of attention recently
with the Planck collaboration publication of a smaller, indirect,
measurement of $H_{0} = 67.3~\pm 1.2\ \mathrm{km\ s^{-1}\ Mpc^{-1}}$
\citep{planck_collaboration_planck_2013} based on modeling the CMB
power spectrum. For a flat $\Lambda$CDM cosmology this constitutes
a $2.4\,\sigma$ tension with the original \shoes{} direct measurement.
This tension is reduced to $1.9\,\sigma$ when including the updates by
\citet{humphreys_correction_2013} and \cite{Efstathiou_H0_2014}.
(See \citealt{bennett_concordance} for a discussion of this tension and its
consequences for cosmology.)

In this paper, we examine the possibility of as-yet-unaccounted-for
environmental dependencies affecting SNe~Ia, 
and the potential for bias in the direct measurement of \Hnot{}. Concerns of potential
environmental biases in standardized SN~Ia distances arise from
both theoretical and empirical studies. A wide range of
progenitor configurations and explosion scenarios remain in contention.
But in all progenitor models, variation is allowed due to differences
in mass, composition, geometrical configuration and evolutionary
stage.  Statistically, the incidence of these factors is modulated
by the parent stellar population, i.e., the progenitor environment.
The theoretical predictions remain far too uncertain to be applied
directly for precision cosmological analyses, motivating empirical
studies of the association between SN properties and environment.

Empirical studies using global or nuclear host-galaxy properties
have been fruitful in revealing environmental dependencies that
remain even after SNe~Ia are standardized using their lightcurve
widths and colors
\citep[e.g.][]{kelly_hubble_2010,sullivan_host_2010,lampeitl_effect_2010,gupta_improved_2011,dandrea_spectroscopic_2011}.  
The clearest relation found in these studies is a ``step'' in the
mean Hubble residual between SNe~Ia in hosts above and below a total
stellar mass of $10^{10.2\pm0.5}\ M_{\sun}$ \citep{childress_sys_2013}.
While there is a predicted trend of SN~Ia luminosity with metallicity
via the effects of neutronization
\citep{hoeflich_1998,timmes_2003,kasen_2009}, this observed change
as a function of host mass via the mass--metallicity relation is
simply too fast. However, there is a strong transition between
predominately passive to predominately star-forming galaxies around
this stellar mass, and a very general star-formation-driven model 
fits the mass step well \citep{childress_sys_2013}.

While such studies based on global host properties have been
productive, they leave unanswered the deeper connection to the
progenitor.  Global measurements of environmental properties are
light-weighted quantities, and thus will be skewed towards the
environmental properties of galaxy cores. 
Slit or fiber spectroscopy
is even more biased in this regard as the outer regions of the
galaxy, in which a SN progenitor may have formed, are either
geometrically deweighted (when integrating along a slit) or excluded
altogether (when using a fiber).
Of course the degree of this bias depends on the ---~generally
unknown~--- projected radial gradients of age and metallicity in
the SN hosts. This can be ameliorated in part by measuring 
host properties in annuli at the same galactocentric radius as the SN
\citep{raskin_2009}.

In \citet[][hereafter R13]{rigault_evidence_2013} we went a step
further and focused on the immediate environment surrounding each
SN.  A key insight that motivated the R13 study was the realization
that the small velocity dispersion of young stars 
\citep[$\sim 3$~km\,s$^{-1}$; ][]{dezeeuw_1999, pz_2010,
roser_2010} means that
the youngest SN~Ia progenitors would not have had time to migrate
from the neighborhood
where they were formed. Thus, if SNe~Ia do indeed have a rapidly
falling ---~$1/t$~--- delay time distribution \citep[see
e.g.][]{maoz_dtd}, then only a minority of SNe~Ia would be
superimposed on a geometrical region of their host that is unrelated
to their birth environment.  Even for such cases, the global
environment would need to be dramatically different than the
progenitor formation environment to produce an incorrect characterization
of environment properties.

Galaxy simulations show only limited radial and azimuthal mixing
in disk galaxies over 1~Gyr timescales, and the coherence over
10~Gyr is still surprisingly good for a large fraction of stars
\citep[see e.g.][]{2008ApJ...675L..65R, 2008ApJ...684L..79R,
2011A&A...534A..75B, 2012MNRAS.420..913B,
2012MNRAS.426.2089R, 2013A&A...553A.102D}. 
For this reason, a local measurement was almost certain
to be superior to a global measurement in terms of isolating
environmental variables influencing progenitor properties.

Using Nearby Supernova Factory observations
\citep[\SNfactory,][]{aldering_overview_2002}, R13 showed that SN~Ia
standardized magnitudes depend on the star formation activity of
the SN environment within a projected radius of 1~kpc, as traced
by \Ha{} surface brightness.  After standardization using SALT2
\citep{guy-salt2-2007}, SNe~Ia in locally passive
environments (designated as \sN) are on average brighter than SNe
in locally star-forming regions (designated as \sS) by 
$\DcHR=$~\SFSNf
\footnote{$\DcHR\equiv\langle~M_B^{corr}~\rangle_{\sS}-\langle M_B^{corr}\rangle_{\sN} $}.
Since the underlying connection is with star
formation rather than the \Ha{} emission itself, we refer to this
effect as the star-formation bias, or SF~bias for short. 

R13 connected the SF~bias to the host-mass step by noting that few
of the \sN{} in the \SNfactory{} sample
occur in low-mass hosts, leading to a shift 
in mean brightness with host
mass that is driven by the changing fraction of star formation. 
However, this also implies that simply correcting for the host-mass
step will not necessarily correct the star-formation bias (see
Appendix~A of R13 for details). As discussed there, since the
fraction of SNe~Ia from passive regions is expected to decrease
with lookback time, such a magnitude difference can introduce a
redshift-dependent bias in distance measurements based on
SNe~Ia. More subtle perhaps is the fact that even variations in the
ratio of passive to star-forming (SF) hosts within nearby SNe~Ia
samples may also induce a bias. This may introduce systematic errors
into peculiar velocity measurements via the star formation~--~density
relation, and could bias the direct measurement of \Hnot{} when the
SN distance ladder relies on distance indicators tied to specific
stellar populations.

Given this, confirmation of the star-formation bias and its impact
on the cosmological parameters ---~notably $w$ and \Hnot{}~--- are of
paramount importance. The bias on $w$ was examined R13; potential
bias on \Hnot{} is a subject of this paper. We split our investigation into two parts.
The first part of the paper, Section~\ref{sec:H09_Ha_bias}, presents
our main analysis confirming the SF~bias in the independent
\constitution{} SN~Ia dataset
compiled in \citet[][hereafter H09]{hicken_constitution_2009}. 
In the second part of the paper, Section~\ref{sec:H_0andH_a},
we investigate how the SF~bias affects
the measurement of \Hnot{} using SNe~Ia. We conclude in
Section~\ref{sec:conclusion}.

\section{Confirmation of a Star Formation Bias}
\label{sec:H09_Ha_bias}

In this part of the paper, we describe the
dataset, measurements and results of our investigation of
the SF~bias using a SN~Ia dataset largely independent of that
used in R13. Section~\ref{sec:H09_R13_subsample} describes the
sample selection, including sources of attrition.
Section~\ref{sec:FUV_LHa_construction} discusses the measurements,
including correction for dust extinction and the choice of local metric
aperture size. Sections \ref{sec:H09_Ha} and \ref{sec:Robustness_H09_Ha}
present our main results regarding confirmation of the SF~bias and the
robustness of the results. In Section~\ref{sec:H09_Bimodality} we discuss
the structure of the Hubble residuals relative to the bimodal model of
R13. Some finer technical aspects of these measurements are given in
Appendices~\ref{sec:MLCS2k2}--\ref{sec:WIM} for the benefit
of interested readers.

\subsection{The Comparison Sample}
\label{sec:H09_R13_subsample}

In order to confirm the SF~bias previously detected in the
\SNfactory{} sample we need an independent nearby Hubble-flow sample for which
it is possible to compare SALT2-standardized magnitudes between
SNe~Ia from locally star-forming and passive environments.  The
compilation of H09 has been used previously for a number of
cosmological analyses \citep[e.g., H09,][]{kessler_cosmo_2009,rest_cosmo_2013,riess_h0_2009,riess_3_2011}
and only six of the SNe~Ia have been studied in R13 already.  For
the nearby Hubble flow range of $0.023<z<0.1$ (as used, e.g.,
by \shoes{}), we find that the H09 compilation contains \NtotalMLCS{} such
Hubble-flow SNe~Ia. Because the H09 sample was constructed for
cosmological applications, peculiar SNe~Ia or those with large
extinction or poor lightcurve fits have already been removed. 
Thus, the H09 compilation appears to be quite
suitable for an independent measurement of the SF~bias, provided a
suitable set of local star-formation measurements can be obtained. 
A $0.14\pm0.07$~mag offset between E/S0 and Sc/Sd/Irr
morphological types in this sample was identified by H09, thus a first
comparison between bias revealed by morphological versus local
star-formation indicators will be possible.

In the case of R13, it was possible to obtain sensitive measurements
of the local \Ha{} surface brightness as a direct by-product of the
SuperNova Integral Field Spectrograph (SNIFS) observations conducted
by the \SNfactory. Conventional
slit spectroscopy or imaging photometry, as employed for the SN~Ia
follow-up programs compiled in H09, does not afford any robust
parallel quantitative measurement of local star formation.
Furthermore, archival \Ha{} imaging is quite limited for the galaxies
in the H09 compilation. Thus, suitable \Ha{} data for measuring the
SF~bias in the H09 sample are not currently available.

The far-ultraviolet (FUV) luminosity is another well-established SF
indicator. Previously, we used this along with optical data to
characterize the global star formation activity of \SNfactory{} 
SNe~Ia host galaxies \citep{childress_sys_2013,childress_data_2013},
while \citet{neill_local_2009} did the same for many host galaxies
in the H09 sample.

This led us to investigate the availability of sufficiently deep
\GALEX{} FUV imaging for the H09 host galaxies.  We found that
\NwithGALEXMLCS{} out of the \NtotalMLCS{} H09 Hubble-flow SN~Ia host
galaxies have UV coverage from the \GALEX{} GR6/7 data release in the MAST
archive\footnote{\url{http://galex.stsci.edu/GR6/}}, and that 
these data have sensitivity sufficient to classify SN~Ia
environments following the scheme of R13 for most hosts.
(Observations from the CAUSE phase of the \GALEX{} mission were not
considered due to their inhomogeneous nature.) 
We also investigated the available coverage from SWIFT. There the
overlap with the H09 Hubble-flow subsample was too small to be useful
and all but two cases had \GALEX{} coverage already, and thus we
decided not to use the SWIFT data at this time.
We therefore proceed to examine the SF~bias using a
combination of the nearby Hubble-flow SN~Ia subset from H09 and
UV data from \GALEX.

As a start, we examine any biases that may arise from excluding
the subset of SNe~Ia lacking \GALEX{} coverage. \GALEX{} was ostensibly
an all-sky imaging survey (AIS) reaching a $5\,\sigma$ point source
depth of $m_{FUV}=19.9$~AB~mag \citep{GALEX_calibration}.  However
partway through the mission the FUV detector failed to function,
leading to incomplete FUV coverage. In addition, bright stars were
avoided in order to prevent damage to the \GALEX{} detectors, leaving
coverage gaps concentrated towards the Galactic plane region, which
SN surveys avoid anyway.  \GALEX{} also conducted deeper surveys 
---~the Medium Imaging Survey (MIS) covering 1000~deg$^2$ to
$m_{FUV}=23.5$~AB~mag and the Deep Imaging Survey (DIS) covering
80~deg$^2$ to $m_{FUV}=25.0$~AB~mag~--- in fields coincident with
other extragalactic surveys. These solid angles are much smaller
than the sky coverage typical of nearby SN surveys, and thus constitute
a fairly random sampling.  Since these variations in \GALEX{} coverage
with respect to sky location or proximity to bright stars are
completely decoupled from characteristics of nearby SN searches,
there is no {\it a priori} expectation for a bias between SNe~Ia in
hosts with and without \GALEX{} coverage.  Indeed, application of
Kolmogorov-Smirnov tests indicates that lightcurve stretch, color
and standardized Hubble residual distributions of the
\NwithGALEXSALT~SNe~Ia 
without \GALEX{} observations are completely compatible with those
having \GALEX{} coverage, giving similarity probabilities
greater than 16\% for all comparisons. 

\input{Table_1_sample_composition}

Next we apply the selection criteria used in R13, which we follow
in order to provide the best possible comparison to that study. 
In R13 we eliminated SNe~Ia spectroscopically classified as 91T-like according
to \cite{scalzo_91T} due to the possibility that they may be so-called
super-Chandrasekhar SNe~Ia and therefore not representative of SN
cosmology samples. In Appendix~\ref{sec:91T} we provide details of
this selection process, which resulted in the elimination of three 91T-like SNe~Ia,
one of which lacked \GALEX{} FUV coverage anyway.  R13 also removed
highly-inclined hosts; as discussed in more detail in
Appendix~\ref{sec:Inclined_hosts}, for FUV observations this helps
avoid both potential false-positive and false-negative environmental
associations.  We identified \NInclinedMLCSword{} SN host
galaxies with $i>80\arcdeg$; SN~1992ag, SN~1995ac, SN~1997dg, SN~1998eg,
SN~2006ak, SN~2006cc and SN~2006gj, and removed
them from our baseline analysis.

As a result of these sample selection procedures, which are summarized
in Table~\ref{tab:H09_sample_composition}, our baseline analysis
will utilize \NmainSALT{} hosts when using Hubble residuals based on
SALT2 and \NmainMLCS{} hosts when using MLCS2k2 \citep{Jha_mlcs} Hubble residuals.
This sample size compares favorably with the sample of 82
hosts used in R13 to discover the SF~bias.

\subsection{Measurement of Local Star Formation}
\label{sec:FUV_LHa_construction}

\subsubsection{FUV  and \Ha{} as Star Formation Indicators}

Massive short-lived O and early B type stars with $\gtrsim 17\ M_\odot$ 
are responsible for the ionizing radiation that generates
\Ha{} emission, while FUV emission is produced by O- through late-B
stars with $\gtrsim 3\ M_\odot$. Detection of UV light is therefore
an indication of star formation within the preceding $100~\mathrm{Myr}$
\citep[see][for a detailed review]{calzetti_sfr_2013} and FUV and
\Ha{} emission are strongly coupled.  This makes them consistent and
commonly used star formation indicators (see e.g.
\citealt{sullivan_SFRUV_2000, bell_kennicutt_2001, salim_uv_2007}
and generally \citealt{lee_SF_2009,lee_SF_2011} and references
therein).

Like \Ha, the FUV flux drops dramatically with the age of the stellar
population. In simple stellar-population instantaneous-burst models
that account for the late-time contribution of hot subdwarfs, the
FUV flux drops by $\sim1.4$~dex between 10~Myr and 100~Myr, and
then another $\sim3$~dex from 100~Myr to 1~Gyr
\citep{han2007_FUV_ssp,leitherer_starburst99_1999}.  While this is
an exceptionally strong signal, it is complicated by dust extinction
that is stronger in the FUV than for \Ha. This not only weakens the
ability to detect star formation, but adds non-negligible uncertainty
arising from the extinction correction. 

There is also a diffuse FUV component, analogous to the diffuse \Ha{} 
commonly observed in nearby spiral galaxies. 
In Appendix~\ref{sec:WIM} we provide further details concerning this
diffuse emission source, along with our examination of its potential
impact. For the equivalent star-formation threshold set in R13 (see
below), we find that this component should only marginally affect our
classification of SN~Ia environments.

After correcting for Galactic and interstellar dust extinction,
details of which are discussed
in Section~\ref{sec:local-dust-correction}, 
both \Ha{} and FUV indicators can be
converted to a star formation rate (SFR) surface density, \SFRSD{} (in $\mathrm{M_\odot\
    yr^{-1}\ kpc^{-2}}$):
  \begin{align}
    \label{eq:FUV_to_LHa}
    \SFRSD &= \kappa_{1} \times {L^0_{\mathrm{FUV}}}/{\mathcal{S}} \\ \nonumber
    \SFRSD &= \kappa_{2} \times \SigmaHa 
  \end{align}
  In this equation, $L^0_{\mathrm{FUV}}$ is the dust-corrected FUV
  luminosity (in $\mathrm{erg\ s^{-1}\ Hz^{-1}}$) summed over
  an aperture centered at the SN location having area $\mathcal{S}\ \mathrm{kpc^{2}}$.
  \SigmaHa{} is the local \Ha{} surface brightness (in $\mathrm{erg\
    s^{-1}\ kpc^{-2}}$).  The redshifts considered here are small
  ($z\sim 0.03$), so the FUV and NUV $K$-corrections are negligible
  ---~typically smaller than the measurement errors
  \citep{2012MNRAS.419.1727C}. 
  Here we use the usual conversion factors $\kappa_{1} = 1.08\times 10^{-28}$ and
  $\kappa_{2} = 5.5\times 10^{-42}$, as in, e.g., \citet[][]{salim_uv_2007,
    kennicutt_09, calzetti_sfr_2013}. Modifications to the initial
mass function, metallicity, etc., can alter these conversion factors by
$\pm0.2$~dex; see Table~2 of \citet{hao_convert_2011} for examples.
As in R13, we do not attempt to perform corrections to face-on
quantities due to uncertainty concerning the 3-dimensional
distribution of star formation in local regions. Even for the extreme case of 
SNe in the planes of pure disks viewed at random inclinations below our limit of $i<80\deg$,
only $\sim0.2$~dex of additional scatter is introduced.

In R13 we used an \Ha{} surface density threshold of $\log(\SigmaHa) =
38.35$~dex, corresponding to $\log({\SFRSD})=-2.9$~dex, to split the
\SNfactory{} sample into two equal-sized groups. Below this threshold SNe~Ia
were classified as having a locally passive environment, \sN, and
above this threshold they were classified as having a locally
star-forming environment, \sS. The R13 threshold also happened to
be that ensuring a minimum $2\,\sigma$ detection over the \SNfactory{} 
redshift range, and it was also high enough to limit the impact of
miscategorization caused by diffuse \Ha{} emission.

We retain this threshold for the current analysis for consistency
with R13. For FUV measurements this threshold is also sufficient
to minimize miscategorization due to the aforementioned diffuse FUV light;
\citet{boquien_threshold_2011} found that for $\log({\SFRSD})>-2.75$~dex
interarm regions in M33 are largely suppressed and our threshold
is only slightly below this. 
The mildly non-linear relation observed between
\Ha{} and FUV \citep{lee_SF_2009,verley_m33_2010} does not affect
the placement of this threshold by more than $\sim 0.1$~dex.

To account for measurement errors, rather than simply dividing
the SNe~Ia into two groups as in R13, we will estimate a probability
for each SN, \pIe{}, giving the chance that its local environment is
locally passive. 
To do so we use the Poisson errors on the measurements of FUV flux and extinction,
$A_{FUV}$ (see Section~\ref{sec:local-dust-correction}), 
and calculate the fraction of the resulting $\log({\SFRSD})$
distribution that has $\SFRSD^{\lim} \leq 10^{-2.9}\ \mathrm{M_\odot\
yr^{-1}\ kpc^{-2}}$.
We have confirmed the appropriateness of using Poisson
uncertainties by measuring aperture fluxes for $10^4$ blank sky
regions and checking that the results were Poisson-distributed.

\subsubsection{Local Dust Correction}
\label{sec:local-dust-correction}

Dust is associated with star formation 
\citep[e.g.,][]{charlot_simple_2000,simones_m31_2014,verley_m33_2010}
and so can have a strong impact on the observed UV light around the
SN location. The amount of dust depends on many factors such as the
geometry, the quantity of metals available to form dust, and dust
production and destruction mechanisms and timescales. Nevertheless,
for star-forming galaxies there is a good correlation between
FUV$-$NUV color and the amount of FUV dust-absorption, $A_{FUV}$.
Here we use the relation given in Eq.~5 of \cite{salim_uv_2007} to
estimate $A_{FUV}$.  (See \citealt{conroy2010} for examples
of several alternative extinction relations.)  Since this correction
is only appropriate for star-forming environments, we face the need
to assess whether an environment is star-forming before knowing
whether to correct for extinction.

We start by examining the global star-forming properties of the SN
host galaxies. The association of dust with star-formation suggests
that locally passive environments should not require extinction
correction, and in R13 we found that globally passive host galaxies
are also locally passive. Thus, in most cases it would be inappropriate
to apply extinction corrections to the local environments for SNe
in globally passive galaxies.  One very useful quantitative measure
of star-formation activity is the global specific star-formation
rate (sSFR).  sSFR measurements are available for $\sim60\%$ of the
host galaxies in our sample.  These are based primarily on UV and
optical colors \citep{neill_local_2009}, plus NIR for some
\citep{childress_data_2013}.  Using sSFR, we categorize host galaxies
with conclusively low sSFR as globally passive and those with
conclusively high sSFR as globally star-forming.  Specifically, to
be considered conclusively low or high, we require that the measured
sSFR be, respectively, one standard deviation below or above a
boundary set at $\sSFR=-10.5$~dex.  In Table~\ref{tab:H09_FUV}, we
designate these as having host types of Pa and SF, respectively.
Cases where the sSFR is within one standard deviation of the
threshold are designated as $\sim$Pa and $\sim$SF, depending
on whether their sSFR is, respectively, below or above $-10.5$~dex.

When a sSFR measurement is not available, we rely on morphology.
\citet{gildepaz_galex_nga_2007} have shown that morphological type
is another useful means of selecting those galaxies that follow the
star-forming UV color relation. Host galaxies with E/S0 morphological
classifications are considered to be globally passive, while later
morphological types are considered to be globally star-forming.
Again, in Table~\ref{tab:H09_FUV} these are designated as Pa or SF,
respectively.

Once these global designations are assigned, we consider
the local environments of SNe in globally passive host galaxies to
be ineligible for extinction correction. That is, those local
environments for SNe in host galaxies with conclusively low sSFR,
or, when sSFR is not available, E/S0 morphology, are not corrected
for extinction.  Those local environments for SNe in host galaxies
that have conclusively high sSFR, or, when sSFR is not available,
non-E/S0 morphology, are corrected using the relation from
\cite{salim_uv_2007}.  For cases with a global sSFR measurement
that is inconclusively passive or star-forming, in accord with the
association of dust with star-formation, we apply an extinction
correction if the local FUV signal is detected at greater than
$2\,\sigma$.  We have checked that using our morphological criterion
in place of sSFR for cases where the sSFR is poorly measured does
not change which cases are corrected for dust.  Finally, the hosts
of SN~2005hc and SN~2005mc were cases of early type galaxies with
inconclusive \sSFR{} measurement where local star-formation was
detected (see Appendix~\ref{sec:bright_core}); extinction corrections
were applied in these cases.

With this procedure we can be fairly certain that an extinction
correction is being applied only when appropriate. Since
the uncertainty on the FUV$-$NUV color is sometimes large, we include
a prior on the resulting $A_{FUV}$ based on the $A_{FUV}$ versus
color distribution measured for spiral galaxies by \cite{salim_uv_2007}.
This prior leads to a typical $A_{FUV}=2.0\pm0.6$~mag for
large FUV$-$NUV uncertainties \citep[see also][]{salim_prior_2005}.
For completeness, we report in Table~\ref{tab:H09_FUV} our best
estimate of $A_{FUV}$ based on FUV$-$NUV color and the
\cite{salim_uv_2007} relation, whether or not it was actually
applied.  With this, the interested reader can examine the impact
of making slightly different choices regarding the extinction
correction. The maximum $A_{FUV}$ allowed in the relation
of \citet{salim_uv_2007} is 3.37~mag; thus extinction correction
can increase $\log({\SFRSD})$ by at most 1.35~dex, and 0.8~dex will
be more typical. Therefore, proper extinction correction is important,
but can only affect the classification of SN hosts whose star-formation
surface density is already near our threshold.

To obtain our final values of \SFRSD, we combine the
uncertainties on the FUV fluxes with the uncertainties on
$A_{FUV}$ by convolving the two probability distribution
functions.  The local extinction-corrected FUV flux is then converted
into \SFRSD{} using Eq.~\ref{eq:FUV_to_LHa}. In
Section~\ref{sec:Robustness_H09_Ha} we explore the effect of applying
a blanket correction to the local environments of all globally
star-forming hosts.

\subsubsection{Local Aperture Size Appropriate for \GALEX\ FUV Data}
\label{sec:Practical_FUV_measurement}

The local star formation measurements in R13 were performed in a
metric aperture of 1~kpc radius; the integrated flux within such a
metric aperture will fade as $1/((1+z)^2\ {d_L}^2)$ for nearby galaxies.
The spatial resolution of SNIFS observations was $\sim1$~arcsecond
FWHM, thus the aperture ranged from 1.2$\times$ to 3.3$\times$ the
spatial resolution over the $0.03<z<0.08$ redshift range of the
\SNfactory\ sample.

While the redshift range of the H09 sample is lower, the 
4\farcs2~FWHM spatial resolution of the \GALEX{}  
FUV channel is considerably worse than that
of the R13 sample observed with SNIFS.
For a point source this means that a
metric aperture of 1~kpc radius will measure a quickly decreasing
fraction of the PSF, resulting in even more signal loss as a function
of redshift. If a galaxy has extended FUV emission, the signal does
not fade in this way, but instead includes more and more contaminating
signal from outside the true local environment as redshift increases.
This could lead to a miscategorization of the local environment,
in particular a star-forming region or diffuse FUV light 
contaminating the signal for a region that is locally
passive. 

As a compromise we have settled on a 2~kpc radius aperture ---~twice
the diameter used in R13.  At the median redshift of our H09 subsample,
$z_{med}=0.032$, this aperture will subtend 6\farcs2 and
enclose approximately 65\% of the FUV light from a compact source
such as an isolated star cluster.  We test the influence of the
aperture size on our results in Section~\ref{sec:Robustness_H09_Ha}.
Then, in Appendix~\ref{sec:dilute_signal} we examine the
extent to which the local environment signal of small hosts might
be diluted due to this larger aperture.

\subsubsection{SN~Ia local FUV Measurements}
\label{sec:Local_FUV_measurement}

The local UV signal is obtained from \GALEX{} images by summing the
number of counts within a 2~kpc radius around the SN location from
the ``\texttt{int}'' images after removing the background signal
given in the ``\texttt{skybg}''
images\footnote{\url{http://www.galex.caltech.edu/researcher/faq.html}}.
Uncertainties arise from photon noise only. Counts are converted
into AB-magnitudes using zero points of $18.82$ and $20.08$ for the
FUV and the NUV channels, respectively \citep{GALEX_calibration}.
The images were inspected for contamination by known AGN or bright
stars; not surprisingly no such cases were found since such
contamination likely would have made a SN a poor choice for cosmology
analyses in the first place (see Appendix~\ref{sec:bright_core}
for the case of a new LINER discovered in this process).

Accurate SN~Ia positions for use in positioning the measurement
apertures are taken from \citet{hicken_H09_2009}, \citet{hamuy1996}
or
NED{\footnote{\url{http://www.ned.ipac.caltech.edu}. We found the
NED coordinates for many SNe~Ia from \citet{hamuy1996} to be in
error.}}.
The astrometric accuracy of the \GALEX\ dataset is 0\farcs59 for FUV and 0\farcs49
for NUV \citep{GALEX_calibration}. The coordinate uncertainties for
the SNe and from \GALEX\ are therefore much smaller than the projected angular
size of our metric aperture.

The measured FUV and NUV fluxes were then corrected for Galactic
extinction using the \citet{schlegel_1998} dust map and the Galactic
extinction curve derived by \citet{cardelli_dust_law} as updated
by \citet{odonnell_dust_law}. For a \citet{cardelli_dust_law} dust
curve parameter of $R_V = 3.1$ this gives $A^{\mathrm{MW}}_{FUV} = 7.9\,E(B-V)$
and $A^{\mathrm{MW}}_{NUV} = 8.0\,E(B-V)$. We assume a
statistical error of 16\% on the values of $A^{\mathrm{MW}}_{FUV}$ and
$A^{\mathrm{MW}}_{NUV}$, correlated between bands \citep{schlegel_1998}.

The resulting measurements are summarized in Table~\ref{tab:H09_FUV}. 

\input{Table_2_main_table}

\subsection{The Star-Formation Bias in the H09 Sample}
\label{sec:H09_Ha}

\begin{figure*}
  \centering
  \includegraphics[width=0.7\linewidth,clip]{%
    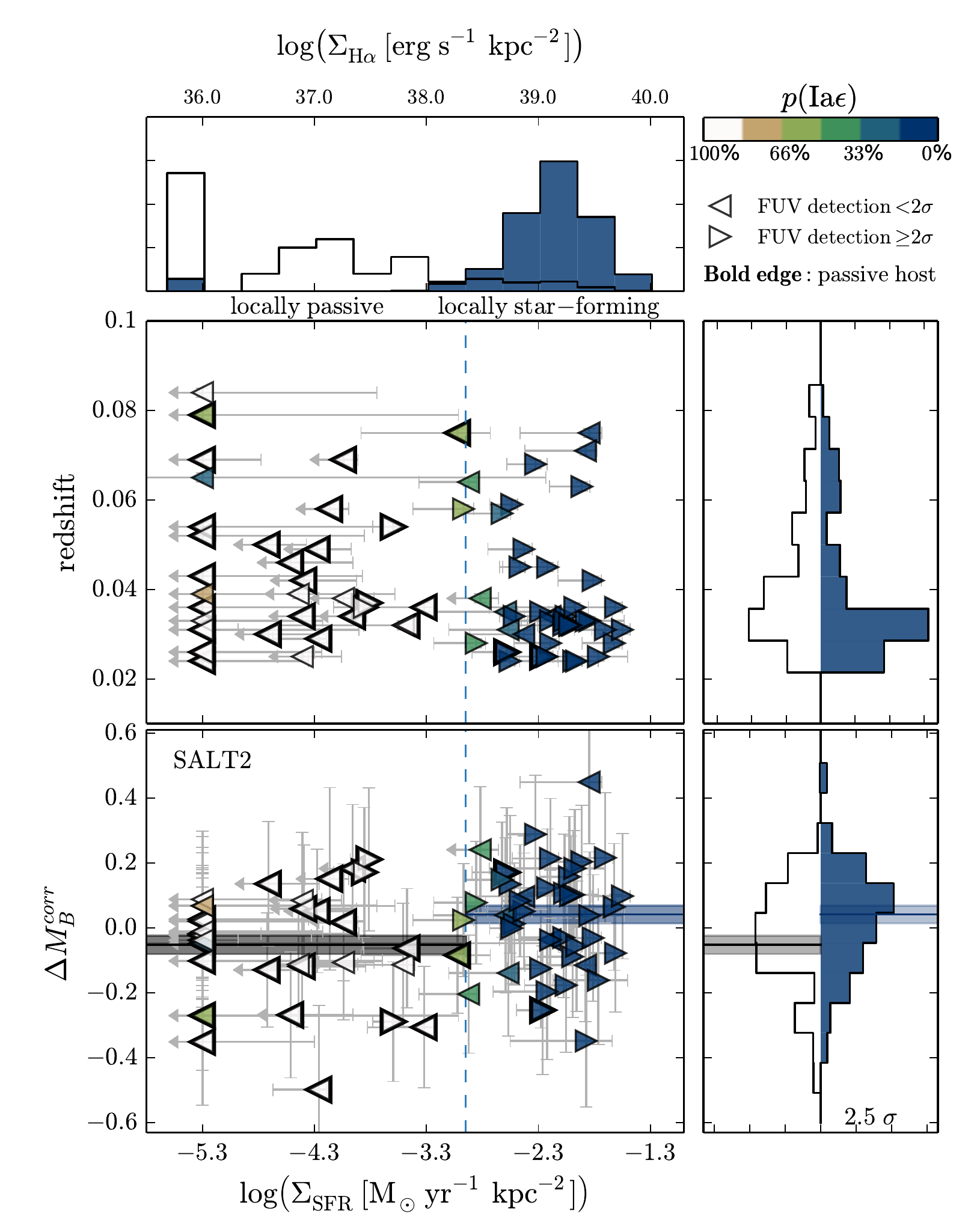}
  \caption{
    SN~Ia redshifts and SALT2 standardized Hubble residuals (\cHR)
    as a function of $\log(\SFRSD)$.  Cases where no counts were
    found in the \GALEX{} FUV image are arbitrarily set to
    $\log(\SFRSD)=-5.3$~dex.  \emph{Upper panel:} the $\log(\SFRSD)$
    distributions for the environmental subgroups.  Each SN contributes
    to the amplitude of the open histogram according to its value
    of \pIe{}, and the filled, green histogram according to its
    value of \pIa{}.  \emph{Main panels:} Marker colors encode the
    value of \pIe{} for each SN.  Those identified as having a
    globally passive host (Pa and $\sim$Pa, as
    defined in Section~\ref{sec:local-dust-correction}), are
    highlighted with thick black marker contours (see the legend
    for details).  The vertical dashed blue line shows our
    $\log(\SFRSD)=-2.9$~dex star-formation surface-density threshold.
    \emph{Right panels, from top to bottom:} Marginal distributions
    of redshift and \cHR{} for each subgroup.  These bi-histograms
    follow the same color code and construction method as the
    \SFRSD{} histograms.  The weighted mean of the \cHR{} values
    of each H09 subsample is drawn over its respective marginal
    distribution in the lower panels.  The transparent bands show
    the $\pm 1\,\sigma$ uncertainty on these means.  Compare to
    Figure~6 of R13, and see Figure~\ref{fig:R13_H09_LHa_bias_MLCS}
    for the MLCS2k2 results.
  }
  \label{fig:R13_H09_LHa_bias}
\end{figure*}

In Figure~\ref{fig:R13_H09_LHa_bias} we show the SALT2-standardized
Hubble residuals, \cHR{}, from H09 as a function of our measurement of
$\log(\SFRSD)$ for the \NmainSALT{} SNe~Ia of the H09/\GALEX{} sample.
We find that the SNe~Ia from locally passive environments are
$\DcHR=$\SFHickenSALT{} brighter. This is the difference between the means
of the Hubble residuals for the two environmental
subsamples, derived from a maximum likelihood calculation
in which each SN has a chance \pIe\ or $1-\pIe$ of belonging to the
\sN\ or \sS\ population, respectively.
The variances on the Hubble residuals from H09
(which includes the intrinsic dispersion they assigned) as listed
in Table~\ref{tab:H09_FUV}, are used in the likelihood calculation.
Because the log-likelihood involves the logarithm of sums unique for
each SN, it must be solved computationally.
The summed probability for the number of \sN{} is \SumsNSALT, representing
\HgalexglobalHFSALT{} of the sample. 

This measurement constitutes an independent confirmation, at the
\SFHickenLEVELSALT{} confidence level, of the SF~bias previously observed
in the \SNfactory{} dataset ($\DcHR=$~\SFSNf; R13).  The amplitude
of the bias is in remarkable agreement between the two samples.  
It is similar to the $0.14\pm0.07$~mag offset 
between E/S0 and Sc/Sd/Irr galaxies found by H09, but has a higher
statistical significance.

We also tested for the presence of this bias in the H09 dataset when
using the MLCS2k2 lightcurve fitter for three commonly used
values of $R_V$, again taking Hubble residuals directly from H09.
H09 found that MLCS2k2 with $R_V=1.7$ produced the smallest dispersion,
and this case gives an observed bias of $\DcHR = $\SFHickenMLCS. 
Similar results are found for $R_V=3.1$ and $R_V=2.5$.
From this we conclude that the SF~bias found in the H09 dataset is
only mildly dependent on which of the two lightcurve fitters
is used, differing by only $\sim1\,\sigma{}$ after
taking into account the covariance between SNe~Ia in common. 
(See \citealt{akim14}
for additional discussion of sensivitity to host environment 
with different lightcurve fitters.)
A deeper knowledge of what is driving the SF bias will
help in understanding such variation between
lightcurve fitters. 
Table~\ref{tab:SFbias_Summary} summarizes these results
(also see Figure~\ref{fig:R13_H09_LHa_bias_MLCS}),
while Figure~\ref{fig:robustness} shows how the results depend on
various analysis choices, as detailed in Section~\ref{sec:Robustness_H09_Ha}.

\input{Table_3_SF_bias_summary}

Finally, for use with Equation~A.14 in Appendix~A of R13, which
details the relation between the star formation bias and the host
mass step, we report the fraction of high mass hosts, $F_H$, in the
H09/\GALEX{} sample.  We find $F_H =$~\HickenGALEXHighMassFracSALT{} for
the H09 sample studied here, and expect it to be representative of
most nearby samples currently in use. This compares with a high-mass
fraction of only $\sim55$\% for the \SNfactory\ \citep{childress_data_2013},
SDSS \citep{gupta_improved_2011} or PTF \citep{pan_host_2014}
samples.

\subsection{Robustness of the H09 Star-Formation Bias}
\label{sec:Robustness_H09_Ha}

\begin{figure*}
  \centering
  \includegraphics[width=1.\linewidth,clip]{%
    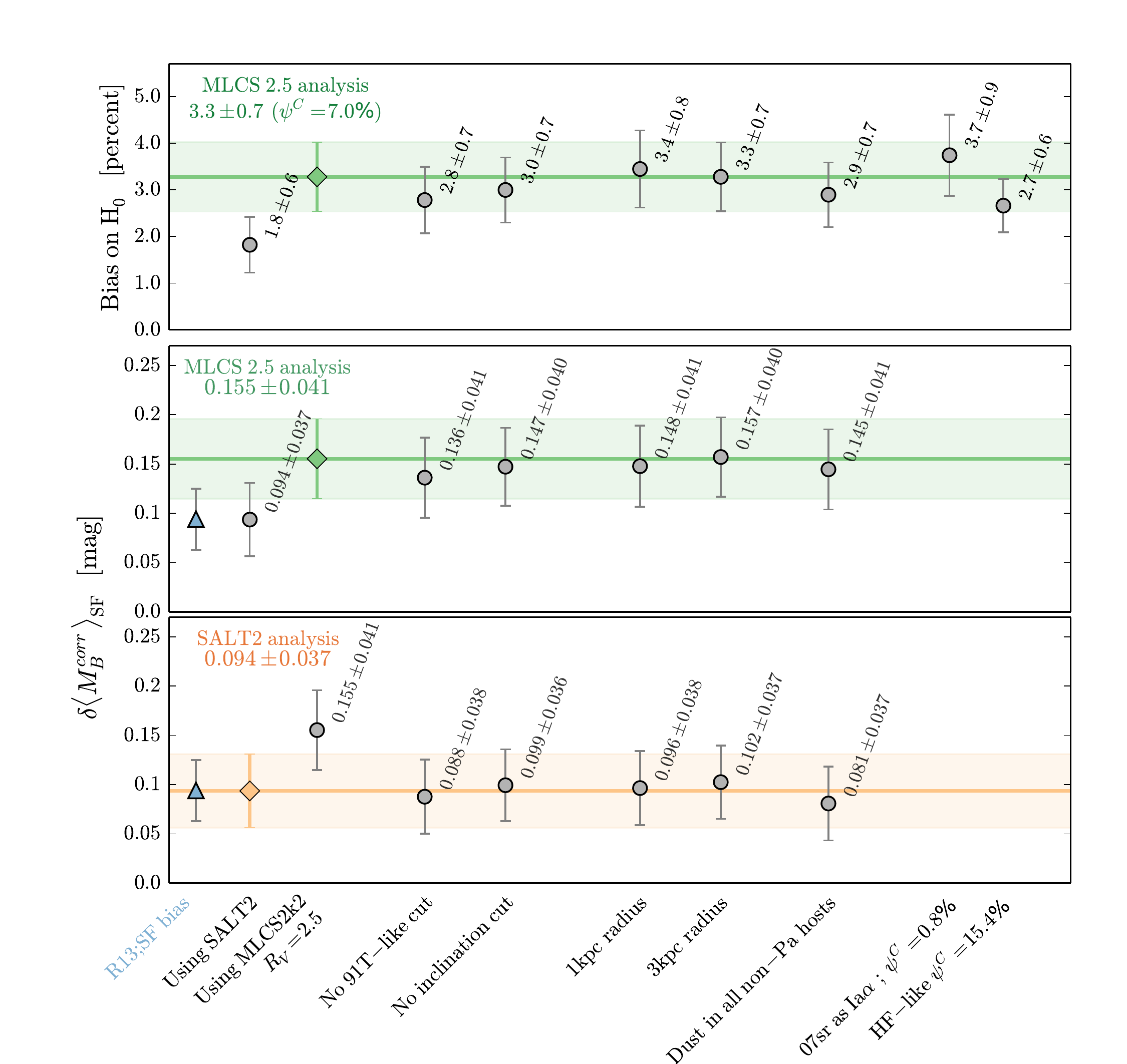}
  \caption{Summary of the influence of the analysis choices made in
    the paper. The main analysis results are indicated in the
    upper left of their corresponding panels 
      and drawn as horizontal lines. The shaded bands indicate 
    the corresponding $\pm1\,\sigma$
    errors.  \emph{Lower and Middle} panels: The SF~bias
    as presented in Section~\ref{sec:H09_Ha_bias}, measured using 
    Hubble residuals from H09 based on SALT2 (lower) and
    MLCS2k2 (middle) lightcurve parameters.  
    \emph{Upper panel}:
    The \Hnot{} bias, as presented in Section~\ref{sec:H_0andH_a},
    using Hubble residuals based on MLCS2k2 lightcurve parameters,
    as in \shoes{}. The main \Hnot\  bias uses $\psi^{C}=$\CepsiMLCS,
    but we also present two variants. We consider the case when 
    SN~2007sr is assumed to be a \sS{}, in which case $\psi^{C}= 0.8\%$,
    as well as the case where the Cepheid hosts are measured with angular resolution
    and signal-to-noise typical of the Hubble-flow sample, in which case
    $\psi^{C}=$\CepsiSimulHRMLCSRiess. The summary results are reported in
    Table~\ref{tab:environmental_H0_bias}.} 
  \label{fig:robustness}
\end{figure*}

In this section, we test the influence of the various criteria used
in performing this portion of the analysis. These include the sample
selection, the radius chosen to represent the local environment and
the dust correction.  The effects of changing each of 
these selections in turn is illustrated in the lower two panels
of Figure~\ref{fig:robustness}, and is discussed in the following
paragraphs. Two key ideas to 
keep in mind here are that a) while the specific values of 
\SFRSD{} can change with details of the measurement technique,
only SNe~Ia near the \SFRSD{} threshold can have much effect, and 
b) any errors made in categorizing the local environment are
most likely to {\it decrease} the measured size of any 
real SF bias by mixing SNe~Ia from different environments.

{\noindent\bf Sample construction:}
{In the main analysis, we made two sample selections (see
Table~\ref{tab:H09_sample_composition}). 
As in R13, we removed 91T-like SNe and the SNe from highly-inclined host galaxies.
Alterations in these selection criteria
} change the SF bias by less than $0.015$~mag, as illustrated in Figure~\ref{fig:robustness}.

{\noindent\bf Local environment measurement radius:}
In Section~\ref{sec:Local_FUV_measurement} we presented the rationale
for our use of a 2~kpc radius aperture to define the ``local'' SN
environments. We have tested the impact of using either a smaller
(1~kpc) or larger (3~kpc) aperture, and find changes of
less than 0.01~mag, regardless of which light-curve fitter Hubble residuals are used.

{\noindent\bf Dust correction:}
In our main analysis, when the global \sSFR{} was not
at least one standard deviation from the threshold set at $-10.5$~dex, 
an extinction correction was applied
only if the FUV signal was detected at
more than $2\,\sigma$ in a host galaxy compatible with being star
forming
(Section~\ref{sec:local-dust-correction}).
Here we test the extreme case of applying dust extinction corrections
to the local environment whenever the host galaxy is not globally passive. 
This tests the possibility that strong
extinction is responsibly for pushing the observed FUV signal below
the $2\,\sigma$ detection threshold.  
This test essentially consists of assigning the
typical $A_{FUV}=2.0 \pm 0.6$~mag found by
\cite{salim_prior_2005,salim_uv_2007} to the SNe~Ia from globally
star-forming hosts that were not corrected for extinction in the
main analysis. This change to the analysis reduces the amplitude
of the SF bias by only $\sim$~0.010--0.013~mag, as shown in
Figure~\ref{fig:robustness}.

{\noindent\bf \GALEX{} sensitivity:}
We have considered whether the limited sensitivity of some \GALEX{}
exposures might affect our results.  Fortunately, almost half of the
hosts considered have exposures several times longer than those of
the main \GALEX{} AIS survey (see Table~\ref{tab:H09_FUV}).  We find
no direct correspondence between SF sensitivity and redshift or
classification probability in the current dataset.

In summary, the measurement of the H09 SF~bias is robust at the
$\sim0.015$~mag level against variations in the analysis.

\subsection{Hubble Residual Bimodality}
\label{sec:H09_Bimodality}

R13 suggested the presence of a bimodal structure in the \cHR{}
distribution of the \SNfactory{} dataset (see the lower-right histogram
in Figure~\ref{fig:R13_H09_LHa_bias} of R13).  The brighter mode
consisted entirely of SNe~Ia from locally passive environments,
\sN, while the fainter mode consisted of SNe~Ia from a mix of \sS{} 
and \sN{} environment. While the fainter mode extended over the full
range of host galaxy masses, the brighter mode was concentrated in
the high-mass half of the distribution.

As in R13 we find that the SNe~Ia having locally star-forming
environments have a $\sim30$\% tighter dispersion than the
overall sample. 
After removing the random noise expected due to small-scale galaxy
peculiar velocities, we find that the \sS{} population has a weighted
RMS of only \IaaDispersionNoPecWithErrorMLCS{} when using
MLCS2k2 Hubble residuals.
See Table~\ref{tab:SFbias_Summary} for a summary of the 
weighted RMS for each environmental subpopulation.

We have fit the R13 bimodal model, without any adjustment, to the
environmentally categorized Hubble residuals of our H09 subset.
Here we find that the R13 model is a better fit than a single
Gaussian whose dispersion is allowed to float.  Specifically, we
measure differences of \AICDeltaAICcMLCS{} and \AICDeltaAICcSALT{} in favor of the bimodal model
for the Akaike information criterion corrected for finite sample
size (AICc) when using MLCS2k2 and SALT2 Hubble residuals, respectively.
The result using MLCS2k2 strongly favors the R13 bimodal model, but
the evidence for bimodality using the H09 SALT2 Hubble residuals is
weaker than found in R13.  The mixed evidence for bimodality here
is not surprising given the generally larger measurement uncertainties
reported by H09.

\subsection{Combined Star Formation Bias}
\label{sec:High_mass_host_combined}

The SF bias measured for the H09 sample is in remarkable agreement
with the value determined in R13 using the \SNfactory{} sample, both
based on SALT2 Hubble residuals.  Under the assumption that the SF
bias is a universal quantity, we can combine these
measurements (first averaging the six SNe~Ia in common).
Doing so, we find a common SF~bias of $\DcHR=$~\SFcombinedSALT{}, which
constitutes a \SFcombinedLEVELSALT{} measurement
of this environmental bias.
Alternatively, this becomes \SFcombinedMLCS{} when combining the
R13 results with the MLCS2k2 Hubble residuals from the H09/\GALEX{}
dataset.  

\section{Consequence of the Star-Formation Bias on the Measurement of \Hnot{}}
\label{sec:H_0andH_a}

In this second half of the paper, we turn to the question of
whether the confirmed SF bias affects the current measurements of \Hnot.
Currently, the most accurate method to directly measure \Hnot{} is
to use SNe~Ia in the Hubble flow (HF) to estimate \Hnot$^2\langle
L_{SN}\rangle$, and then calibrate the average standardized SN~Ia
luminosity, $\langle L_{SN}\rangle$, using SNe~Ia with accurate and
independent distance measurements. The period---luminosity relation
of a large number of Cepheid variable stars within a modest number
of nearby SN~Ia host galaxies has been used to provide a calibration
of $\langle L_{SN}\rangle$ \citep{riess_3_2011, freedman_2012}.
Underlying this approach is the assumption that $\langle L_{SN}\rangle$
is the same in both the Cepheid-calibrated and the Hubble-flow
samples.

However, the Cepheid-calibrated sample targets globally young
environments since Cepheids are very young stars, with ages less
than 100~Myr. Indeed, the main sequence counterparts of classical
Cepheids are B-type stars like those contributing to the FUV flux
we use here as a star formation indicator \citep[for a review
see][]{turner_cepheid_review}.  Thus, the fraction of SNe~Ia in
star-forming environments (\sS) is likely to be higher for the
Cepheid sample than for the Hubble flow sample.  The average SN~Ia
standardized luminosity will then differ between the two samples
due to the SF~bias demonstrated in Section~\ref{sec:H09_Ha}; this, in
turn, biases the measurement of \Hnot{}. 
In the following sections, we estimate the amplitude of this bias on
the Hubble constant. 

The correction to \Hnot{}, resulting in an unbiased measurement,
\Hnotcorr{}, can be written quite generally as:
\begin{align}
  \label{eq:H0_Ha_bias}
  \log(\Hnotcorr) = \log(H_0) 
  - \underbrace{
    \frac{1}{5}(\psi^{HF}-\psi^{C})\times \DcHR.}_{\text{SF~bias correction}}
\end{align}
Here $\psi^{C}$ and $\psi^{HF}$ respectively denote the fraction
of SNe~Ia in the {\it specific} Cepheid-calibrated and Hubble-flow
samples being compared which have locally passive (\sN) environments.
The two terms on the right-hand side work together; even
if there is a SF bias, it only biases \Hnot{} if $\psi^C$ and
$\psi^{HF}$ are not equal. The appropriate values of $\psi$ are
calculated by taking the average of the \pIe\ probabilities for the
corresponding datasets. Conceptually, the net effect of
Eq.~\ref{eq:H0_Ha_bias} is to form the weighted average of two
Hubble diagrams ---~one for SNe~\sS\ and one for SNe~\sN.

Assuming that the SF~bias, \DcHR{}, is a universal quantity, it may
be determined from the specific sample under study, or by including
external measurements. Note that the value of $\psi$ for any external
sample used solely to measure \DcHR{} is immaterial in this context.
At most it affects the sensitivity of the \DcHR{} measurement from
the external sample; it does not enter into $\psi^{HF}$.
Conversely, if there are SNe~Ia used to calculate both the
original \Hnot{} measurement and \DcHR{}, as is the case here, there
will be positively correlated errors between these two quantities. 

Having established the value of \DcHR{} in Section~\ref{sec:H09_Ha}, we
now evaluate the other inputs to Eq.~\ref{eq:H0_Ha_bias}, the
fraction of SNe~Ia with locally passive environments in the
Hubble flow and Cepheid-calibrated SN~Ia host galaxies.
Because the main analysis in \shoes{} used MLCS2k2 $R_V=2.5$ Hubble
residuals, we do so here.

\subsection{The Fraction of \sN{} Among the \shoes{} Cepheid Galaxies}
\label{sec:Cepheid_psiCp}

We first examine the eight SNe~Ia hosts whose distances were measured
using the Cepheid period--luminosity relation by \shoes{}.
We measure their \pIe{} in the same manner as for the
H09 sample, as described in Section~\ref{sec:H09_R13_subsample}.
The results are summarized in Table~\ref{tab:SHOES}.  The top half
of the table summarizes the galaxies used by \shoes{} to measure
\Hnot, while the bottom half presents our measurements for additional
SN~Ia host galaxies whose Cepheid-based distances are anticipated.
The local environments for seven of the eight are covered by both
\GALEX{} FUV and NUV observations, while SN~1998eq lacks FUV coverage.

Although SN~1998aq does not have FUV imaging, it does have a strong
NUV signal.  We can use the NUV signal along with very conservative
assumptions to categorize the local environment of SN~1998eq as star
forming.  Specifically, even assuming an extreme UV color of
FUV$-$NUV$=1$ \citep[see Figure~13 of][]{salim_uv_2007} and no dust
extinction still requires a minimum value for the local star formation
density of $\log(\SFRSD)>-2.46$~dex.

SN~2007sr is an exceptional case, as it is located in the well-known
tidal tail of the merging galaxies NGC~4038/39 (the Antennae).
Tidal environments are known as sites of strong star formation
\citep[e.g.][]{neff_tidal_2005,smith_tidal_2010,kaviraj_tidal_2012}.  The
Antennae tidal tail is indeed quite blue \citep{hibbard_blue_tail_2005},
and contains star clusters with mean ages of 10--100~Myr
\citep{whitmore_cluster_ages_1999,fall_cluster_ages_2005}.  Therefore,
one might presume that SN~2007sr should be classified as a \sS.
However, SN~2007sr actually lies at a projected separation of 2~kpc
from the spine of the tidal tail, and \citet{hibbard_blue_tail_2005}
estimate an age around 400~Myr along the portion of the tail projected
closest to SN~2007sr. (Note that the \citet{hibbard_blue_tail_2005}
age determinations did not include correction for possible dust
extinction, and so may be too large.)  An age of 400~Myr is older
than the estimated dynamical time for the tidal tail, and would
therefore suggest that stars near the location of SN~2007sr were
originally formed in the disk of their parent galaxy.  
Tidal forces will act in a similar
manner on the volume of stars originally local to the SN~2007sr
progenitor, but it is likely that the stars are now more spread out
due to dynamical evolution, having moved SN~2007sr off the spine of
the tail and lowering the stellar surface density, and thus
the measured \SFRSD, compared to the
original environment.

When faced with such ambiguity for the Hubble-flow sample, as
with highly-inclined host galaxies,  
the simplest avenue was to cut them from the sample. Doing
that in this case would result in a value $\psi^{C}\sim0$, along with
a slight increase in the uncertainty on \Hnot{} due to the smaller
number of calibrators, since all the remaining environments for the
Cepheid-calibrated SNe~Ia are star-forming.  Given the possibility
that SN~2007sr may be slightly too old to be counted as \sS, this
choice could slightly overestimate the final correction to \Hnot{} 
that is needed. To better reflect the ambiguity for this calibrator,
we will take a neutral value of $\pIe=0.5$ for our main analysis.
We note that SN~2005cf, listed in Table~\ref{tab:SHOES} as a likely
future calibrator, is also located in a tidal tail.  Thus, the
question of the proper categorization of tidal tails will resurface
when it is time to incorporate SN~2005cf into the measurement of
\Hnot.

Combining the eight individual \pIe{}, we find $\psi^{C}=$\CepsiMLCS{} 
as the fraction of SNe~\sN{} in the Cepheid-calibrated SNe~Ia used
for the \shoes{} measurement of the local Hubble constant.  As
anticipated, the local environments of this sample is predominately
star forming.

\input{Table_4_Cepheid_table}

\subsection{ The Fraction of \sN{} in our H09/\GALEX\ Hubble-Flow Sample}

If we had estimates of \pIe{} for all 140 nearby Hubble-flow SNe~Ia used by
\shoes{} we could calculate and apply the resulting $\psi^{HF}$
directly in Eq.~\ref{eq:H0_Ha_bias}. However, $\sim30$ of the
Hubble-flow SNe~Ia used by \shoes{} are not contained in H09, and
in addition, we do not have \pIe{} estimates for another
30 due to insufficient \GALEX{} coverage or selection cuts 
(see Table~\ref{tab:H09_sample_composition}).
Thus, unlike the above estimate of
$\psi^{C}$, our estimate of $\psi^{HF}$ will need to be statistical.
 
The fact that we still have $\sim60$\% of the SNe~Ia in common will
reduce the uncertainty substantially since we incur a statistical
error (beyond that already incorporated in the \pIe{} values for
individual SNe) only for the $\sim40$\% that remain unmeasured.
We have shown in Section~\ref{sec:H09_R13_subsample} that the
H09/\GALEX{} dataset is statistically representative of the entire H09 sample,
and we expect that to be true for the $\sim30$ \shoes{} Hubble-flow
SNe~Ia not in H09. 

For the H09/\GALEX{} subset having MLCS2K2 measurements we
find $\psi^{HF}=$~\HgalexglobalHFMLCSRiess.
For those with SALT2 measurements
$\psi^{HF}=$~\HgalexglobalHFSALT.  This is in agreement with the
value $\psi^{HF}=$~\HSNFHF{} in R13.  More comparable
is the subset of high-mass hosts in R13, for which the value is
slightly higher, $\psi^{HF}=$~\HSNFhighmassHF. This may reflect
differences between the untargeted \SNfactory{}  search that provided
most SNe~Ia in R13 and galaxy-magnitude-limited searches that
provided most of the SNe~Ia in H09. It also could be due to a greater
chance of false-positive associations due to the $4\times$ greater
area covered by the larger aperture used here.  
Applying
$\psi^{HF}=$~\HgalexglobalHFMLCSRiess{} as a statistical estimator to the
59 \shoes{} SNe~Ia environments we were unable to measure, 
and combining with those we have measured, gives $\psi^{HF}=$~\HFpsiMLCSRiess.
The variations due to the alternatives explored in
Section~\ref{sec:Robustness_H09_Ha} are within the final statistical
uncertainty on $\psi^{HF}$ for the \shoes{} Hubble-flow sample, and
are reflected in the calculations of the variations shown in
Figure~\ref{fig:robustness}.

\input{Table_5_H0_environmental_bias}

\subsection{Hubble Constant Corrected for the Star-Formation Bias}
\label{sec:new_Hnot}

As a result of the high fraction of local star-forming environments for the
SNe~Ia in the \shoes{} sample, they do not provide a calibration that
is representative of the H09 nearby Hubble-flow sample.  In
Section~\ref{sec:Robustness_H0} we will present further insights
into, and explore the robustness of this measured difference in
the values of $\psi^{HR}$ and $\psi^{C}$.

Using Eq.~\ref{eq:H0_Ha_bias} with values $\psi^{HF}=$~\HFpsiMLCSRiess,
$\psi^{C}=$\CepsiMLCSRiess, and $\DcHR=$~\HFDcHRMLCSRiess{} we estimate that the
Hubble constant is currently overestimated by \HnotbiasMLCSRiess{} due
to the SF~bias. As will be discussed in
Section~\ref{sec:SF_bias_Mass_step}, a small fraction of this bias
most likely has been taken into account already because \shoes{} applied
a 0.75\% correction to account for potential host-mass dependency in the
SN standardized magnitudes. Since the correction for the
star-formation bias also corrects for the host-mass effect 
(see Section~\ref{sec:SF_bias_Mass_step}), we
estimate an effective correction to \Hnot{} of
$-$\EffBiasPercentEftaTMLCSRiessMassRmv{} (see
Table~\ref{tab:environmental_H0_bias}).

There have been several updates to the basic ingredients since the
\shoes{} analysis. \citet{humphreys_correction_2013} reported an
improved value for the distance to the NGC~4258 megamaser that served
as one of the zero-points for the Cepheid distance scale (along
with the LMC distance and parallaxes for Milky Way Cepheids).
\citet{Efstathiou_H0_2014} re-examined the selection of Cepheids and
the resulting calibration of the Cepheid period--luminosity relation
for the SN host galaxies, slightly modifying those distances.  Since
it is the most recent, and includes the revised distance to NGC~4258,
we will use the \citet{Efstathiou_H0_2014} analysis as our baseline
in quantifying how the SF~bias affects the apparent tension
between direct and indirect measurements of \Hnot. Note that
correction for the SF~bias involves a change to the luminosities
assigned to SNe~Ia in the Hubble flow relative to those in the
Cepheid sample; we make no adjustments to the distances assigned
to the Cepheid-calibrated SN~Ia host galaxies.

By applying Eq.~\ref{eq:H0_Ha_bias}, we calculate a value of the
Hubble constant corrected for the SF bias of
$\Hnotcorr$=\HnotCorrEftaTMLCSRiessMassRmv{} when using the LMC distance,
Milky Way parallaxes and the NGC~4258 megamaser as the Cepheid
zeropoint as in \shoes{} and \citet{Efstathiou_H0_2014}. If the
NGC~4258 megamaser is used as the sole zeropoint of the Cepheid
distances and the new \citet{humphreys_correction_2013} distance
is used with the \citet{Efstathiou_H0_2014} analysis, the revised
\Hnot{} is lower, but less certain, at \HnotCorrEftaOMLCSRiessMassRmv.
Table~\ref{tab:H0_corr} summarizes the SF-bias corrections
for both the \shoes{} and \cite{Efstathiou_H0_2014} estimations of \Hnot{}, 
when using either the megamaser or the three Cepheid anchor(s). 
Details of the contributions to each correction are
given, along with the new level of agreement with several
recent indirect CMB-based measurements of \Hnot. 
We emphasize that the \shoes{} analysis and our corrections
both use MLCS2k2 with R$_V=2.5$, and thus are fully consistent with 
regard to choice of lightcurve fitter.
Our revised \Hnot{} is now compatible at the $1\,\sigma$ level with
these indirect measurement of the Hubble constant for a flat
$\Lambda$CDM cosmology.

We note that this estimate can be improved in the future
by the various teams who employ SNe~Ia to measure \Hnot, by improved
matching of the local SN environments between the calibrator and
Hubble-flow samples.

\input{Table_6_H0_corr_table}

\subsection{Robustness of the \Hnot\ Correction}
\label{sec:Robustness_H0}

The robustness of our correction to \Hnot{} depends on the robustness
of the inputs to Eq.~\ref{eq:H0_Ha_bias}. In
Section~\ref{sec:Robustness_H09_Ha} we have already explored the
robustness of the star-formation bias to changes in our sample
selection and measurement procedures.  The other key
ingredient for the bias on \Hnot{} is the difference in the values
of $\psi^{HF}$ and $\psi^{C}$.  In the upper panel of
Figure~\ref{fig:robustness}, we show how these analysis
parameters affect the measurement of the \Hnot{} bias (see
Section~\ref{sec:Robustness_H09_Ha}).  None have a significant
impact.

{\noindent\bf SN~2007sr classification:}
To further explore the difference between $\psi^{HF}$ and $\psi^{C}$
we consider the case where SN~2007sr is \sS{}. This increases the
SF~bias on \Hnot{} by $0.4\%$; hence if one considers this a
false-negative due to tidal dilution of the local star-formation
surface density, then an extra $-0.3\ \mathrm{km\ s^{-1}\ Mpc^{-1}}$
should be applied to the aforementioned revised \Hnot{} values. The
signs of these changes reverse if SN~2007sr is instead considered
\sN. These changes are within our quoted uncertainty.

{\noindent\bf Simulation of distance effects:}
We next considered possible effects due to the redshift difference
between the two samples by simulating what value of $\psi^{C}$ would
have resulted if the Cepheid-calibrated hosts had been observed
with the same distances and exposures as the H09 Hubble-flow sample.
For each of the six \sS{} Cepheid-calibrated hosts having direct FUV
detections, the star-formation surface densities were measured using
the full set of procedures described in Section~\ref{sec:H09_Ha_bias}
after degrading the spatial resolution and increasing the noise to
match each of the Hubble-flow galaxies.  The net effect, averaged
over all pairings, was to increase $\psi^{C}$ to \CepsiSimulHRMLCSRiess. 
This was primarily due to changes in \pIe{} for the host of
SN~1990N when simulated at higher redshift or with shorter \GALEX{} exposure.
The resulting change in the \Hnot{} bias is once again well within the
quoted uncertainties. This is due in part to the fact that we are
already accounting for noise by using the full \pIe\ probability
distribution functions.

{\noindent\bf Chance of low $\psi^{C}$:}
One might expect the Cepheid-calibrated sample to reflect the \sN{} 
fraction of \FracglobalSFlocallyPMLCSRiess\ found for globally star-forming
galaxies (as defined in Section~\ref{sec:local-dust-correction})
in the Hubble-flow sample. In fact the \sN{} fractions are consistent;
application of Fisher's Exact
Test\footnote{http://en.wikipedia.org/wiki/Fisher's\_exact\_test},
applicable to two sets of categorical data, as here, gives a
\FisherPvalueZeroIaeEightCepheidsMLCSRiess{} chance of finding no locally
passive environments for the \shoes{} sample.

Good consistency is also found when comparing based solely on
morphology.
The SNe~Ia hosts with Cepheid calibration have Hubble types Sb---Sm.
In our Hubble-flow sample we find that for SN host galaxies of these
types the passive fraction is $\sim 26$\%.  Application of
Fisher's Exact Test for this case shows that the passive fraction
in the Hubble-flow and current Cepheid-calibrated samples have a
17\% probability of being the same. Even including the low
locally-passive fraction for the Cepheid-calibrated SNe expected
to be used in the future (see Table~\ref{tab:SHOES}), the probability
of a common locally-passive fraction is 14\%.

{\noindent\bf Consistent, low \cHR\ dispersions:}
One final indicator that the SNe~Ia in Cepheid hosts are of class
\sS{} is that they share the small dispersion that R13 found for
this subclass.
Using Table~3 of \shoes{} we calculated an unbiased 
weighted dispersion in Cepheid-calibrated standardized SN~Ia absolute
magnitudes of only $0.123\pm0.034$~mag.  After removing the dispersion
due to Cepheid distance measurement errors, the SN~Ia dispersion
drops to only 0.101~mag. This agrees well with the dispersion of
$0.099\pm0.009$~mag found by R13 for their SNe~Ia associated with
star-forming environments, and with the
\IaaDispersionNoPecWithErrorMLCSRiess{} 
dispersion of the \sS{} in the H09 Hubble-flow sample (see
  Table~\ref{tab:SFbias_Summary}).  The small 
dispersion of the Cepheid-calibrated SNe~Ia and its similarity to
the dispersion of the R13 and H09 \sS{} subsets could be a coincidence,
but as the overall sample dispersion is much larger,
\IaDispersionNoPecWithErrorMLCSRiess, the likelihood ratio strongly
favors our result that these SNe~Ia truly belong to the \sS{} category.

\subsection{Connection to the Host Mass Step Bias Correction}
\label{sec:SF_bias_Mass_step}

\shoes{} argued that the step in Hubble residuals occurring around
a host mass of $10^{10}\ M_{\sun}$ has a small impact on their estimation of \Hnot{}
since the host masses for the Cepheid-calibrated SNe~Ia are similar
to those of the Hubble-flow SNe~Ia from H09.  Using a linear
correction of Hubble residual versus host mass based on
external information, the \shoes{} estimate
for \Hnot{} is reduced by only 0.75\% (see their Section~3).

Here we have shown that even for a sample such as that of H09 where
almost all hosts are on the high-mass side of the mass step, there
is a bias that is just as large as past studies have found for the
size of the host-mass correction over the full mass range of SN~Ia host
galaxies \citep[e.g.,][]{childress_sys_2013}. 
These results and those of R13 indicate that
star formation is a more important driver than host mass, and that
in fact, the host-mass step may simply be a projection of the star
formation bias in combination with the rapid change in the fraction
of star-forming galaxies as a function of mass as in Figure~11 of
\citet{childress_sys_2013}.

If the SF~bias is indeed the more deep-rooted cause of the mass~step,
then correcting for the SF~bias naturally corrects for the host-mass
dependency (see also Appendix~A of R13). The H09 sample was used
by \citet{kelly_hubble_2010} to originally discover the SN~Ia host
mass bias; for our subset --- having significant overlap with that
of \citet{kelly_hubble_2010} --- we find that the mass step calculated
at their division point of $10^{10.8}\ M_{\sun}$ drops from
$0.100\pm0.040$~mag before correction for the SF bias to
$0.026\pm0.039$~mag afterward. This demonstrates that the SF~bias
does in fact remove the mass step.
Consequently, for our analysis to be self-consistent we
remove the 0.75\% offset to \Hnot{} applied by \shoes{}, where the
intent was to account for the host-mass effect, since our SF~bias
correction has removed it from the data.

\section{Conclusion}
\label{sec:conclusion}

We have used the nearby SNe~Ia from the independent \constitution{} 
dataset compiled by H09 to confirm the local environment bias in
SNe~Ia standardized magnitudes first seen by R13 in the Nearby Supernova Factory
SN~Ia sample. Using Hubble residuals as presented by H09 along with
our \GALEX-based measurements of the local star formation, we confirm this bias
for standardization using either
SALT2 or MLCS2k2 lightcurve parameters with $\DcHR=$\SFHickenSALT{} and
$\DcHR=$\SFHickenMLCSRiess, respectively.  Whereas R13 used \Ha{} as
star formation indicator, here we use the FUV flux as 
measured by \GALEX. Together these results demonstrate that the
effect is not particular to the sample, lightcurve standardization,
or type of star formation indicator that is used.

Combining our new SALT2 bias measurement with that from R13 we find that
SNe~Ia from locally star-forming environments are
$\DcHR=$~\SFcombinedSALT{} fainter after standardization than those
from locally passive environments. This constitutes a
\SFcombinedLEVELSALT{} measurement of
this environmental bias. We also measure that in nearby SN~Ia samples
dominated by high-mass host galaxies, such as that drawn from
\constitution, typically $\sim50$\% of the SNe~Ia arise in locally
passive environments. 

This star formation bias has a direct consequence on the precision
measurement of \Hnot{}, as exemplified by the \shoes{} program.
We find the local environments of SNe~Ia whose absolute magnitudes
were calibrated by \shoes{} using the Cepheid period--luminosity
relation are overwhelmingly star-forming. This in itself is not
surprising because Cepheids are young stars. However, in the presence
of a star formation bias, it means that the standardized magnitudes
of these SNe~Ia will be dimmer on average than those of the Hubble-flow
comparison sample. By applying our measurements of the amplitude of
the star formation bias along with the relative fractions of SNe~Ia in
star-forming or passive local environments for each of the
Cepheid-calibrated and Hubble-flow samples, we find that a
\HnotbiasMLCSRiess{} correction
($-$\HnotBiasAmplEftaTMLCSRiessMassRmvNoUnit~km~s$^{-1}$~Mpc$^{-1}$)
to the \shoes{} measurement of
\Hnot{} is required. The final corrected value, including the revised
NGC~4258 megamaser distance \citep{humphreys_correction_2013}, the
refined Cepheid analysis of \cite{Efstathiou_H0_2014}, and 
backing out the small host-mass correction implemented in \shoes,
becomes $H_{0}=$~\HnotCorrEftaTMLCSRiessMassRmv{}
when starting from the main \shoes{} analysis using MLCS2k2 SN
lightcurve parameters.  
This corrected value for the Hubble constant is within $1\,\sigma$\ of the 
current CMB indirect estimations of the Hubble constant from
Planck \citep{planck_collaboration_planck_2013,spergel_New_Planck}
and WMAP \citep{bennett_WMAP9, hinshaw_WMAP9}. We note
that while our corrected value for \Hnot\ lies within the
uncertainties quoted by \shoes, it is a $\sim5\times$ larger correction
than their quoted uncertainty on the zeropoint from SNe~Ia in
the Hubble flow.

Looking forward to future
studies to measure \Hnot, we present star-formation measurements
for additional nearby SNe~Ia whose distances may soon be calibrated
using the Cepheid period--luminosity relation. These are also locally
star forming, and so the star formation bias will need to be taken
into account in future analyses. Further, we note that
similar caution is necessary when using Type~Ia SNe with other distance 
indicators, such as the Tully-Fisher relation,
surface brightness fluctuations and tip of the red giant branch,
which target galaxies where star formation is unusually high or low.

We also note the discovery of new star formation at SN~Ia sites in
several morphological E/S0 galaxies; such cases may lead to a better
understanding of the connection between different star formation
histories and the properties of SNe~Ia.

\acknowledgements{
We thank Jake Simones, Evan Skillman, Adam Riess, Robert Kirshner,
and Saurahb Jha for useful discussions.  This work is based on
observations made with the NASA Galaxy Evolution Explorer. \GALEX{} 
is operated for NASA by the California Institute of Technology under
NASA contract NAS5-98034.  This work was supported in part by the
Director, Office of Science, Office of High Energy Physics, of the
U.S. Department of Energy under Contract No. DE-AC02- 05CH11231;
in Germany by the DFG through TRR33 ``The Dark Universe;'' in France
by support from CNRS/IN2P3, CNRS/INSU, and PNC, and in China from
Tsinghua University 985 grant and NSFC grant No~11173017.  LPNHE
acknowledges support from LABEX ILP, supported by French state funds
managed by the ANR within the Investissements d'Avenir programme
under reference ANR-11-IDEX-0004-02. NC is grateful to the LABEX Lyon
Institute of Origins (ANR-10-LABX-0066) of the Universit\'e de Lyon for
its financial support within the program ``Investissements d'Avenir''
(ANR-11-IDEX-0007) of the French government operated by the National
Research Agency (ANR).  This 
research has made use of the NASA/IPAC Extragalactic Database (NED),
which is operated by the Jet Propulsion Laboratory, California
Institute of Technology, under contract with the National Aeronautics
and Space Administration.  Some of the data presented in this paper
were obtained from the Mikulski Archive for Space Telescopes (MAST).
STScI is operated by the Association of Universities for Research
in Astronomy, Inc., under NASA contract NAS5-26555. Support for
MAST for non-HST data is provided by the NASA Office of Space Science
via grant NNX13AC07G and by other grants and contracts.
}


\clearpage

\begin{appendix}

\renewcommand\thefigure{\thesection.\arabic{figure}}    
\section{Illustration of MLCS2k2-based Hubble Residuals}
\label{sec:MLCS2k2}
\setcounter{figure}{0}   

\begin{figure*}
  \centering
  \includegraphics[width=0.7\linewidth,clip]{%
    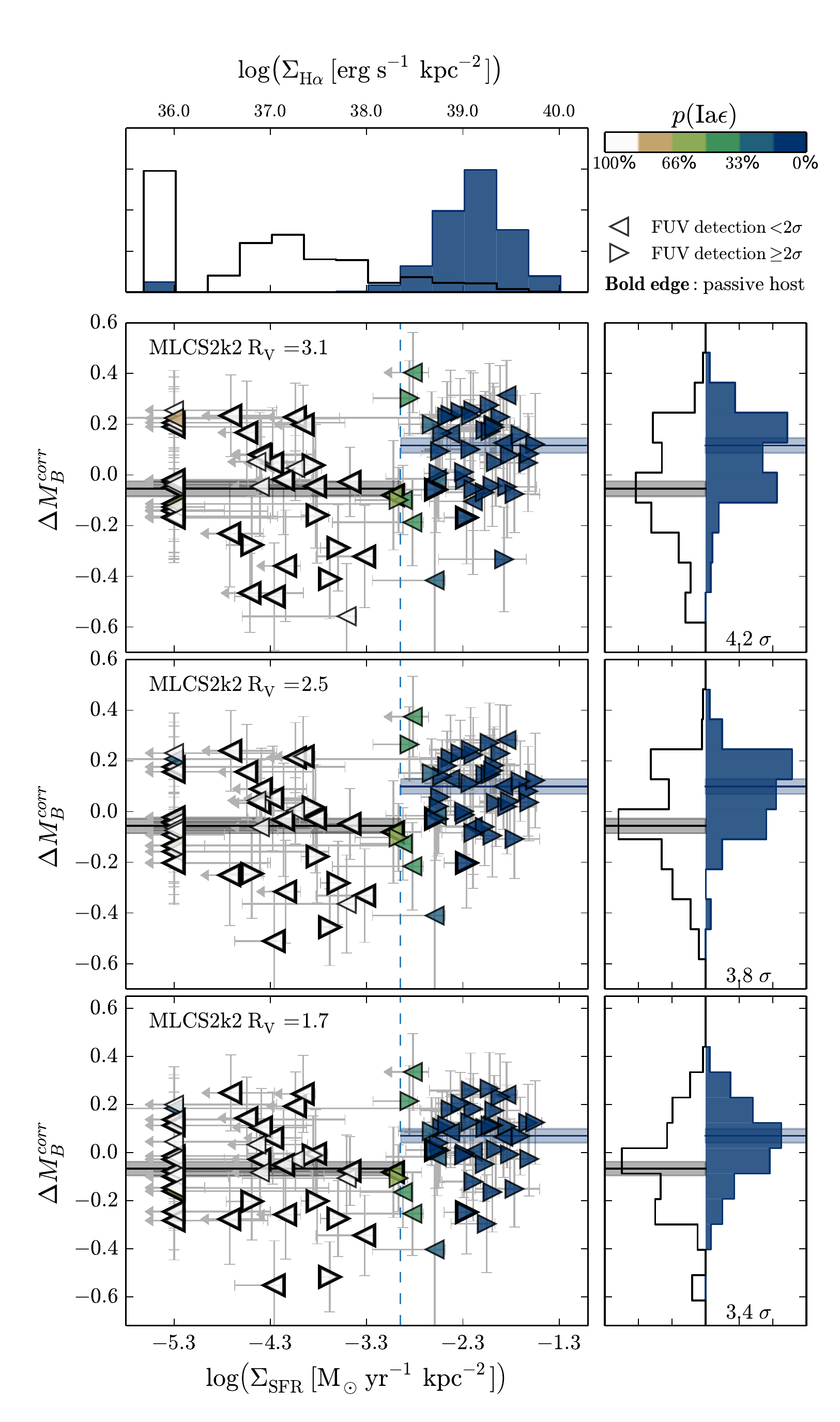}
  \caption{
    SN~Ia MLCS2k2 standardized
    Hubble residuals (\cHR) as a function of $\log(\SFRSD)$,
    \emph{from the bottom to the top:} $R_V$ of 1.7, 2.5 and 3.1,
    respectively. 
    This plot is constructed like Figure~\ref{fig:R13_H09_LHa_bias}.
    }
  \label{fig:R13_H09_LHa_bias_MLCS}
\end{figure*}

The characteristics of the \sS{} and \sN{} SN~Ia subsets shown in 
Figure~\ref{fig:R13_H09_LHa_bias} used Hubble residuals derived
using lightcurve fit parameters determined using SALT2. In
Figure~\ref{fig:R13_H09_LHa_bias_MLCS} we present the analogous
results using Hubble residuals based on MLCS2k2 lightcurve fit
parameters.

\section{R13 Sample Cuts}
\label{sec:R13-cuts}

\subsection{Potential Super-Chandrasekhar SNe}
\label{sec:91T}

In \cite{scalzo_91T}, we suggested that 91T-like SNe~Ia are good
candidates for having masses above the Chandrasekhar limit, and
demonstrated that some indeed do. Because of the possibility that
these might result from a different progenitor channel, they were
not considered in R13. \citet{scalzo_91T} determined the rate of
pure 91T-like SNe~Ia ---~excluding those similar to the less extreme
99aa-like subclass~--- to be $2.6^{+1.9}_{-0.7}$\% of a total SN
rate for a volume-limited sample. Thus, we expect 91T-like SNe~Ia
to be a minor component of the H09 sample.  \citet{Blondin_specto_2012}
have spectroscopically classified most SNe Ia in the H09 sample,
however, they did not distinguish between 91T-like and 99aa-like
events. \citet{silverman_classification} present spectroscopic
classifications for a large sample of nearby SNe Ia having significant
overlap with the H09 sample, and they do distinguish between 91T-like
and 99aa-like events. Combining those studies, we therefore
exclude SN~1998ab and SN~2003hu from our baseline analysis on the
basis of their 91T-like character. SN~1999gp is also 91T-like,
but has no \GALEX{} data or MLCS2k2 measurement. The resulting
91T-like fraction of 2.7\% matches the result of \cite{scalzo_91T}.
Therefore, in this regard the parent population of the SNe~Ia sample
studied here is in agreement with that of R13.

\subsection{Highly Inclined SN Host Galaxies}
\label{sec:Inclined_hosts}
In R13, highly-inclined galaxies were identified as cases where
a locally passive environment might incorrectly be associated with
an \Ha{} signal due to projection.
In the present case, however, highly 
inclined galaxies are as likely to suffer suppressed FUV emission due to
strong dust absorption \citep[see, e.g.,][]{conroy2010}. Thus, it
is unclear whether inclination is more likely to produce false-positive
or false-negative environment categorizations.

Within the set of host galaxies studied here, SN~1992ag,
SN~1995ac, SN~1997dg, SN~2006ak and SN~2006cc 
are highly inclined and are classified as \sS, and therefore may
be examples where a locally passive environment has been miscategorized
as a star-forming environment due to projection. On the other hand,
SN~1998eg and SN~2006gj are examples of SNe~Ia in edge-on galaxies with
no FUV signal, which might be suppressed due to dust extinction.  Following R13,
these cases are removed from our main analysis. In
Section~\ref{sec:Robustness_H09_Ha} we examined whether including
them in the main sample has any effect on our results, and found it
did not. We note that such edge-on galaxies would not be suitable for
Cepheid measurements either.

\section{Additional Potential Cases of Miscategorization}
\label{sec:false-positive}
\subsection{Bright Elliptical Cores}
\label{sec:bright_core}
Another type of miscategorization can occur when 
the FUV signal does not
directly trace star formation, namely when the signal originating from
the very inner cores of E/S0 galaxies surpasses our chosen threshold simply
due to the large column of stars along our line of sight
\citep[e.g. see FUV surface brightness profiles for E/S0 galaxies
in][]{marino_Ecore_2011}. While this problem is far worse in the NUV
band, we considered it prudent to examine this in FUV as well. SNe Ia are rarely
detected on E/S0 galaxy cores due to low contrast, subtraction errors,
and the overall small fraction of stars projected onto the core
relative to the entire galaxy, so only a dozen cases needed this
special attention.

We identified two cases of SNe close to the core of E/S0 galaxies
for which the \GALEX{} FUV signal is above our threshold: SN~2005hc and
SN~2005mc. These have \sSFR{} measurements \citep{neill_local_2009}
which are ambiguous. In the case of SN~2005hc,
there is an apparent star-forming disk present,
which in optical light is hidden by the glare of old stars, and the SN is projected
onto this disk. Therefore, the categorization of the local environment
of SN~2005hc as \sS{} seems correct.  We note that in this case a
categorization based on morphology alone ---~which is commonly
applied in SN host studies~--- would produce erroneous results.

In the case of SN~2005mc, located in (R')SB0 galaxy UGC~4414
\citep{rc3}, there is FUV light at the very core, as well as FUV
light from a known star-forming ring. The SN lies close to the core,
which is slightly bluer in FUV$-$NUV than the cores of normal E/S0
galaxies but not as blue as the star-forming ring. The SDSS spectrum
of the core reveals evidence for LINER activity, with stellar-absorption-corrected 
\Ha{} emission significantly weaker
than [NII]~$\lambda 6583$~\AA\ or [SII]~$\lambda~6717,6731$~\AA\
\citep{bpt,veilleux_AGN_1987,kewley_AGN_2006}.  Thus, the signal
in the core may be due in part to the high column density of old stars
projected onto the core combined with LINER activity. To test this
we remeasured the FUV flux in a 1~kpc aperture, which safely clears
any light from the core. The resulting \SFRSD{} did not change
significantly, therefore, we retain classification of SN~2005mc as
having a \sS{} environment.

As part of this study we examined SN~2003ic, which has an
elliptical host galaxy but where \GALEX{} shows signs of recent
star-formation. It's observed \SFRSD, which was not corrected for
extinction since the host appeared to be globally passive, is below
our threshold and gives $\pIe=100\%$.  If an extinction correction
had been applied, its \SFRSD{} would have been slightly above our
threshold, with $\pIe\sim30\%$.  It is thus likely to be the third
case of a SN~Ia associated with comparatively recent star-formation
hidden in an otherwise normal-looking elliptical host galaxy.
\citet{forster_SFinETG_2008} and \citet{schawinsky_SFinETG_2009}
have also searched for such cases. This SN~Ia is discussed extensively
in \citet{kelly_hubble_2010} since it is in the largest host galaxy
in their sample and thus helps drive some of the trends they report.

\subsection{Diluted FUV Signal from Star-Forming Environments}
\label{sec:dilute_signal}

Here we consider the possibility that an actively star-forming
environment may fall below our threshold due low stellar surface
density or due to geometrical dilution. Potential examples 
of these situations may include low surface brightness galaxies,
tidal tails, dwarf galaxies, or an aperture extending
beyond the nominal edge of the SN host galaxy.

Because the vast majority of SNe~Ia in the H09 sample came from
searches that targeted galaxies selected from magnitude-limited
catalogs, the sample contains few dwarf galaxies and no known SNe~Ia
in tidal tails or giant low surface brightness galaxies.  Nor is
the drop in surface brightness with galactocentric radius a major
consideration for the galaxies considered here; inspection of the
\GALEX{} Ultraviolet Atlas of Nearby Galaxies \citep{gildepaz_galex_nga_2007}
shows that for galaxies typical of those hosting SNe~Ia in the H09
sample, when there is active SF it rises above our SF surface density
threshold over most of the face of the galaxy.  Indeed, within the
canonical $D_{25}$ optical diameter \citep{rc3}, the
azimuthally-averaged FUV surface brightness
range is typically only around $\sim1$~dex \citep{gildepaz_galex_nga_2007}.
Much of this variation is due to the change in filling fraction 
with radius rather than change in local surface brightness.
This compares with to the 3--4~dex amplitude of the FUV surface density
signal.

However, we have identified the hosts of SN~1999aw and SN~2006an
as dwarf galaxies whose sizes are smaller than the \GALEX{} PSF.
They were both found in large-area surveys unbiased towards SNe~Ia
in known galaxies.  We estimate colors of $B-V\sim0.4\pm0.3$ for
the hosts of SN~1999aw and SN~2006an, based on photometry from
\citet{strolger_99aw_2002} and SDSS, respectively.  While these
optical colors are not nearly as blue as those of the 125~Myr old
host of SN~2007if \citep{childress_sn2007if_2011}, the colors for
the hosts of SN~1999aw and SN~2006an are consistent with those of
the blue compact dwarfs presented in \citet{hunter_dwarfs_2010},
all of would have high star formation surface densities given FUV
observations of comparable resolution and sensitivity.

In SDSS, the host of SN~2006an has a Petrosian radius of
$2.4\pm0.5$~arcsec, or $1.6\pm0.3$~kpc at its redshift of $z=0.065$.
Thus, the effect of geometrical dilution should be small. Still,
because its size is comparable to the \GALEX{} PSF, it will be
suppressed by as much as 0.3~dex since our aperture will not encompass
all of the flux of a compact star-forming region close to SN~2006an.
Thus, our current limit of $\log(\SFRSD) < -2.2$~dex could increase
to as much as $\log(\SFRSD) < -1.9$~dex.

\citet{strolger_99aw_2002} find that the host of SN~1999aw is barely
spatially resolved, having ${\rm FWHM}\sim0.4$~arcsec, or $\sim0.1$~kpc
at its redshift $z=0.038$. Thus the geometrical dilution would be
$\sim0.6$~dex for a 2~kpc radius aperture relative to the 1~kpc
aperture used in R13. Measurement of the FUV flux from the host of
SN~1999aw also will be suppressed by roughly 0.3~dex because our
aperture doesn't cover the full PSF and the galaxy is unresolved
by \GALEX.  Combining both effects, our current limit of 
$\log(\SFRSD) < -3.5$~dex could increase to roughly $\log(\SFRSD) < -2.6$~dex. 
However, if the area over which to measure SF were limited to the
size of this tiny galaxy, that would raise the limit to something
more like $\log(\SFRSD) < -0.5$~dex.

For both dwarfs, much deeper FUV data would be needed in order to
turn our current upper limits into detections able to firmly classify
these two dwarfs. Since the current analysis has only two such cases,
their impact is negligable. However, this may become an issue given
the higher fraction of SNe~Ia in dwarf galaxies now being found in
large-area surveys unbiased towards known galaxies.

Besides the geometrical dilution for small galaxies, there will be
cases where part of our UV measurement aperture extends beyond the
nominal edge of the SN host galaxy. The region outside the host
will bias \SFRSD{} low, though there is likely to be some compensation
from inner, and often brighter, host light spread outwards by the
\GALEX{} PSF. However, it is difficult to define an actual galaxy
``edge,''; at UV wavelengths disks often extend 20--40\% beyond the nominal
$D_{25}$ optical diameter \citep{barnes_uvdisk_2011}. Therefore,
we have not implemented any compensation for this potential
bias.

\section{Diffuse FUV Emission}
\label{sec:WIM}
In star-forming galaxies there 
is a diffuse FUV component that is comparable to the
diffuse \Ha{} observed outside HII regions.
In the case of \Ha{} it remains uncertain whether this diffuse
component arises directly from recombination in the warm 
interstellar medium or scattering off dust particles in the
interstellar medium. Further information on the characteristics
of the diffuse \Ha{} can be found in
\citet{oey_survey_2007}, \citet{seon_witt_2012}
and references therein, and in Section~3 of R13.
\citet{diffuse_FUV_Galaxy} compare the diffuse
\Ha{} and FUV for the Galaxy, finding them to be highly correlated.
They estimate that for high latitudes 37\% of the \Ha{} in the Galaxy
is in a diffuse component on average, but with wide variation. They
estimate that in the Galaxy the direct and diffuse components are
comparable. \citet{diffuse_FUV_M33} find that 65\% of FUV surface
brightness measured on a scale of 1.5~kpc arise from a diffuse 
component in M33. This compares with a 45\% diffuse component for
\Ha.  We have applied their procedure to a number of spiral galaxies
in the \GALEX{} Nearby Galaxy Atlas \citep{gildepaz_galex_nga_2007},
but instead using a 2~kpc radius aperture (to match what we
use for the SN host measurements), and find diffuse FUV
fractions in the range 30--80\%.  Such a diffuse component may
increase the chance that a SN~Ia with a passive local
environment will be mistakenly characterized as a star-forming
environment due to projection onto strong diffuse FUV emission.
However, even if the diffuse FUV emission is due to scattering,
it is mostly likely to be associated with regions of active
star formation over the 2~kpc scale used here.
Fortunately, the star formation surface density spans such a large
range ---~1000$\times$ in this study and in R13~--- that these
factors do not contribute much uncertainty to star formation
categorizations for the majority of SNe~Ia.

\end{appendix}

\end{document}

%% file: Number_H0paper.tex
\newcommand{\SFSNf}{$0.094 \pm 0.031$~mag}
\newcommand{\HSNFhighmassHF}{$63.4\pm7.5$\%}
\newcommand{\HSNFHF}{$50.0\pm5.3$\%}
\newcommand{\NtotalSALT}{104}
\newcommand{\NmainSALT}{77}
\newcommand{\NTlikeSALT}{3}
\newcommand{\NInclinedSALT}{7}
\newcommand{\NInclinedMLCSword}{seven}

\newcommand{\NnoGALEXSALT}{18}
\newcommand{\NwithGALEXSALT}{86}

\newcommand{\SumsNSALT}{38}

\newcommand{\HgalexglobalHFSALT}{\ensuremath{48.9}\%}
\newcommand{\HickenGALEXHighMassFracSALT}{\ensuremath{89}\%}

\newcommand{\SFHickenSALT}{\ensuremath{0.094 \pm 0.037\ \mathrm{mag}}}

\newcommand{\SFcombinedSALT}{\ensuremath{0.094 \pm 0.025\ \mathrm{mag}}}
\newcommand{\SFHickenNomagSALT}{\ensuremath{0.094 \pm 0.037}}

\newcommand{\SFHickenLEVELSALT}{\ensuremath{2.5\,\sigma}}
\newcommand{\SFcombinedLEVELSALT}{\ensuremath{3.8\,\sigma}}

\newcommand{\HnotbiasSALT}{\ensuremath{1.8 \pm 0.6}\%}

\newcommand{\IaDispersionNoPecWithErrorNomagSALT}{\ensuremath{0.147 \pm 0.019}}
\newcommand{\IaaDispersionNoPecWithErrorNomagSALT}{\ensuremath{0.144 \pm 0.024}}
\newcommand{\IaeDispersionNoPecWithErrorNomagSALT}{\ensuremath{0.144 \pm 0.026}}

\newcommand{\EffBiasPercentRiessSALTMassRmv}{$1.1\pm0.6$\%}

\newcommand{\AICDeltaAICcSALT}{$-0.9$}

\newcommand{\NtotalMLCS}{110}
\newcommand{\NmainMLCS}{83}
\newcommand{\NTlikeMLCS}{2}
\newcommand{\NInclinedMLCS}{7}

\newcommand{\NnoGALEXMLCS}{18}
\newcommand{\NwithGALEXMLCS}{92}

\newcommand{\HgalexglobalHFMLCS}{\ensuremath{52.2}\%}

\newcommand{\SFHickenMLCS}{\ensuremath{0.136 \pm 0.040\ \mathrm{mag}}}

\newcommand{\SFcombinedMLCS}{\ensuremath{0.110 \pm 0.024\ \mathrm{mag}}}
\newcommand{\SFHickenNomagMLCS}{\ensuremath{0.136 \pm 0.040}}

\newcommand{\SFHickenLEVELMLCS}{\ensuremath{3.4\,\sigma}}

\newcommand{\HnotbiasMLCS}{\ensuremath{2.9 \pm 0.7}\%}

\newcommand{\CepsiMLCS}{\ensuremath{7.0}\%}

\newcommand{\IaaDispersionNoPecWithErrorMLCS}{\ensuremath{0.127 \pm 0.019\ \mathrm{mag}}}

\newcommand{\IaDispersionNoPecWithErrorNomagMLCS}{\ensuremath{0.165 \pm 0.011}}
\newcommand{\IaaDispersionNoPecWithErrorNomagMLCS}{\ensuremath{0.127 \pm 0.019}}
\newcommand{\IaeDispersionNoPecWithErrorNomagMLCS}{\ensuremath{0.180 \pm 0.014}}

\newcommand{\EffBiasPercentRiessMLCSMassRmv}{$2.1\pm0.7$\%}

\newcommand{\AICDeltaAICcMLCS}{$-4.6$}

\newcommand{\NtotalMLCSRiess}{105}
\newcommand{\NmainMLCSRiess}{81}
\newcommand{\NTlikeMLCSRiess}{2}
\newcommand{\NInclinedMLCSRiess}{7}

\newcommand{\NnoGALEXMLCSRiess}{15}

\newcommand{\HgalexglobalHFMLCSRiess}{\ensuremath{52.1}\%}

\newcommand{\FracglobalSFlocallyPMLCSRiess}{\ensuremath{24.9}\%}
\newcommand{\SFHickenMLCSRiess}{\ensuremath{0.155 \pm 0.041\ \mathrm{mag}}}

\newcommand{\SFHickenNomagMLCSRiess}{\ensuremath{0.155 \pm 0.041}}

\newcommand{\SFHickenLEVELMLCSRiess}{\ensuremath{3.8\,\sigma}}

\newcommand{\HFDcHRMLCSRiess}{\ensuremath{0.155 \pm 0.041\ \mathrm{mag}}}
\newcommand{\HnotbiasMLCSRiess}{\ensuremath{3.3 \pm 0.7}\%}

\newcommand{\CepsiMLCSRiess}{\ensuremath{7.0}\%}
\newcommand{\CepsiSimulHRMLCSRiess}{\ensuremath{15.4}\%}

\newcommand{\HFpsiMLCSRiess}{\ensuremath{52.1 \pm 2.3}\%}

\newcommand{\IaDispersionNoPecWithErrorMLCSRiess}{\ensuremath{0.166 \pm 0.012\ \mathrm{mag}}}
\newcommand{\IaaDispersionNoPecWithErrorMLCSRiess}{\ensuremath{0.122 \pm 0.020\ \mathrm{mag}}}

\newcommand{\IaDispersionNoPecWithErrorNomagMLCSRiess}{\ensuremath{0.166 \pm 0.012}}
\newcommand{\IaaDispersionNoPecWithErrorNomagMLCSRiess}{\ensuremath{0.122 \pm 0.020}}
\newcommand{\IaeDispersionNoPecWithErrorNomagMLCSRiess}{\ensuremath{0.181 \pm 0.015}}

\newcommand{\FisherPvalueZeroIaeEightCepheidsMLCSRiess}{$18$\%}

\newcommand{\HnotCorrHumphMLCSRiessMassRmvNoUnit}{$70.2\pm3.0$}

\newcommand{\HnotBiasAmplHumphMLCSRiessMassRmvNoUnit}{$2.4\pm0.5$}
\newcommand{\CMBTensionHumphMLCSRiessMassRmv}{$0.9\sigma\ /\ 0.3\sigma\ /\ 0.7\sigma$}

\newcommand{\MassBiasAmplHumphMLCSRiessMassRmvNoUnit}{$0.5$}

\newcommand{\HnotCorrEftaTMLCSRiessMassRmv}{$70.6\pm2.6\ \mathrm{km\ s^{-1}\ Mpc^{-1}}$}
\newcommand{\HnotCorrEftaTMLCSRiessMassRmvNoUnit}{$70.6\pm2.6$}

\newcommand{\HnotBiasAmplEftaTMLCSRiessMassRmvNoUnit}{$2.4\pm0.5$}
\newcommand{\CMBTensionEftaTMLCSRiessMassRmv}{$1.2\sigma\ /\ 0.5\sigma\ /\ 1.0\sigma$}
\newcommand{\EffBiasPercentEftaTMLCSRiessMassRmv}{$2.6\pm0.7$\%}

\newcommand{\MassBiasAmplEftaTMLCSRiessMassRmvNoUnit}{$0.5$}

\newcommand{\HnotCorrEftaOMLCSRiessMassRmv}{$68.8\pm3.3\ \mathrm{km\ s^{-1}\ Mpc^{-1}}$}
\newcommand{\HnotCorrEftaOMLCSRiessMassRmvNoUnit}{$68.8\pm3.3$}

\newcommand{\HnotBiasAmplEftaOMLCSRiessMassRmvNoUnit}{$2.3\pm0.5$}
\newcommand{\CMBTensionEftaOMLCSRiessMassRmv}{$0.5\sigma\ /\ 0.2\sigma\ /\ 0.2\sigma$}

\newcommand{\MassBiasAmplEftaOMLCSRiessMassRmvNoUnit}{$0.5$}

\newcommand{\HnotCorrRiessMLCSRiessMassRmvNoUnit}{$70.8\pm2.5$}

\newcommand{\HnotBiasAmplRiessMLCSRiessMassRmvNoUnit}{$2.4\pm0.5$}
\newcommand{\CMBTensionRiessMLCSRiessMassRmv}{$1.3\sigma\ /\ 0.6\sigma\ /\ 1.1\sigma$}
\newcommand{\EffBiasPercentRiessMLCSRiessMassRmv}{$2.6\pm0.7$\%}

\newcommand{\MassBiasAmplRiessMLCSRiessMassRmvNoUnit}{$0.5$}

\newcommand{\NtotalMLCSMW}{109}
\newcommand{\NmainMLCSMW}{84}
\newcommand{\NTlikeMLCSMW}{3}
\newcommand{\NInclinedMLCSMW}{7}

\newcommand{\NnoGALEXMLCSMW}{16}

\newcommand{\HgalexglobalHFMLCSMW}{\ensuremath{52.4}\%}

\newcommand{\SFHickenNomagMLCSMW}{\ensuremath{0.171 \pm 0.040}}

\newcommand{\SFHickenLEVELMLCSMW}{\ensuremath{4.2\,\sigma}}

\newcommand{\HnotbiasMLCSMW}{\ensuremath{3.6 \pm 0.7}\%}

\newcommand{\IaDispersionNoPecWithErrorNomagMLCSMW}{\ensuremath{0.183 \pm 0.011}}
\newcommand{\IaaDispersionNoPecWithErrorNomagMLCSMW}{\ensuremath{0.137 \pm 0.018}}
\newcommand{\IaeDispersionNoPecWithErrorNomagMLCSMW}{\ensuremath{0.202 \pm 0.013}}

\newcommand{\EffBiasPercentRiessMLCSMWMassRmv}{$2.9\pm0.7$\%}



%% file: authors.tex
\newcommand{\BERLIN}{1}
\newcommand{\LBNL}{2}
\newcommand{\IPNL}{3}
\newcommand{\LPNHE}{4}
\newcommand{\UWASH}{5}
\newcommand{\YALE}{6}
\newcommand{\THCA}{7}
\newcommand{\CPPM}{8}
\newcommand{\BONN}{9}
\newcommand{\CLERMONT}{10}
\newcommand{\CRAL}{11}
\newcommand{\UCB}{12}
\newcommand{\FSU}{13}
\newcommand{\IPMU}{14}
\newcommand{\WEAVER}{15}

\author
{
    M.~Rigault\altaffilmark{\BERLIN},     
    G.~Aldering\altaffilmark{\LBNL},
    M.~Kowalski\altaffilmark{\BERLIN},
    Y.~Copin\altaffilmark{\IPNL},  \\
    P.~Antilogus\altaffilmark{\LPNHE},
    C.~Aragon\altaffilmark{\LBNL,\UWASH},
    S.~Bailey\altaffilmark{\LBNL},
    C.~Baltay\altaffilmark{\YALE},
    D.~Baugh\altaffilmark{\THCA} ,
    S.~Bongard\altaffilmark{\LPNHE},  \\
    K.~Boone\altaffilmark{\LBNL,\UCB},
    C.~Buton\altaffilmark{\IPNL},
    J.~Chen\altaffilmark{\THCA}, 
    N.~Chotard\altaffilmark{\IPNL},
    H.~K. Fakhouri\altaffilmark{\LBNL},
    U.~Feindt\altaffilmark{\BERLIN,\BONN}, \\
    P.~Fagrelius\altaffilmark{\LBNL,\UCB},
    M.~Fleury\altaffilmark{\LPNHE}, 
    D.~Fouchez\altaffilmark{\CPPM},  
    E.~Gangler\altaffilmark{\CLERMONT},
    B.~Hayden\altaffilmark{\LBNL},
    A.~G.~Kim\altaffilmark{\LBNL}, \\
    P.-F.~Leget\altaffilmark{\CLERMONT},
    S.~Lombardo\altaffilmark{\BERLIN},
    J.~Nordin\altaffilmark{\BERLIN,\LBNL},
    R.~Pain\altaffilmark{\LPNHE},
    E.~Pecontal\altaffilmark{\CRAL},
    R.~Pereira\altaffilmark{\IPNL},  \\
    S.~Perlmutter\altaffilmark{\LBNL,\UCB},
    D.~Rabinowitz\altaffilmark{\YALE},
    K.~Runge\altaffilmark{\LBNL},
    D.~Rubin\altaffilmark{\LBNL,\FSU},
    C.~Saunders\altaffilmark{\LBNL}, \\
    G.~Smadja\altaffilmark{\IPNL},
    C.~Sofiatti\altaffilmark{\LBNL,\UCB}, 
    N.~Suzuki\altaffilmark{\LBNL,\IPMU},  
    C.~Tao\altaffilmark{\THCA,\CPPM},
    B.~A.~Weaver \altaffilmark{\WEAVER}
}
             
\altaffiltext{\BERLIN}
{
    Institut f\"ur Physik, Newtonstr. 15, 12489 Berlin, 
    Humboldt-Universit\"at zu Berlin
}
\altaffiltext{\LBNL}
{
    Physics Division, Lawrence Berkeley National Laboratory, 
    One Cyclotron Road, Berkeley, CA, 94720
}
\altaffiltext{\IPNL}
{
    Universit\'e de Lyon, F-69622, Lyon, France ; Universit\'e de Lyon 1, Villeurbanne ; 
    CNRS/IN2P3, Institut de Physique Nucl\'eaire de Lyon.
}
\altaffiltext{\LPNHE}
{
    Laboratoire de Physique Nucl\'eaire et des Hautes \'Energies,
    Universit\'e Pierre et Marie Curie, Universit\'e Paris Diderot Paris, CNRS/IN2P3, 
    4 place Jussieu, 75005 Paris, France
}
\altaffiltext{\UWASH}
{
    College of Engineering, University of Washington
    Seattle, WA, 98195
}
\altaffiltext{\YALE}
{
    Department of Physics, Yale University, 
    New Haven, CT, 06250-8121
}
\altaffiltext{\THCA}
{
    Tsinghua Center for Astrophysics, Tsinghua University, Beijing 100084, China 
}
\altaffiltext{\CPPM}
{
    Centre de Physique des Particules de Marseille , Aix-Marseille Universit\'e , CNRS/IN2P3, 163, avenue de Luminy - Case 902 - 13288 Marseille Cedex 09, France
}
\altaffiltext{\BONN}
{
    Physikalisches Institut, Universit\"at Bonn,
    Nu\ss allee 12, 53115 Bonn, Germany
}
\altaffiltext{\CLERMONT}
{
    Laboratoire de Physique Corpusculaire de Clermont-Ferrand, France
}
\altaffiltext{\CRAL}
{
    Centre de Recherche Astronomique de Lyon, Universit\'e Lyon 1,
    9 Avenue Charles Andr\'e, 69561 Saint Genis Laval Cedex, France
}
\altaffiltext{\UCB}
{
    Department of Physics, University of California Berkeley,
    366 LeConte Hall MC 7300, Berkeley, CA, 94720-7300
}
\altaffiltext{\FSU}
{
  Department of Physics, Florida State University, Tallahassee, FL 32306, USA
}
\altaffiltext{\IPMU}
{
   Kavli Institute for the Physics and Mathematics of the Universe, 
   University of Tokyo, Kashiwa 277-8583, Japan
}
\altaffiltext{\WEAVER}
{
    Center for Cosmology and Particle Physics,
    New York University,
    4 Washington Place, New York, NY 10003, USA
}

%% file: Table_1_sample_composition.tex
\begin{deluxetable}{lcccc}
  \centering
  \tabletypesize{\small}
  \tablecolumns{5}
  \tablewidth{0pt}
  \tablecaption{Composition of the comparison sample.
    \label{tab:H09_sample_composition}}
  \tablehead{ & \multicolumn{4}{c} { Number of SNe~Ia}\\ 
                     & {SALT2} & \multicolumn{3}{c}{  MLCS2k2}\\
                     &&{\footnotesize$R_V=1.7$} & {\footnotesize$R_V=2.5$}& {\footnotesize$R_V=3.1$}
  }
  \startdata
  H09 sample within $0.023<z<0.1$ & \NtotalSALT&\NtotalMLCS&\NtotalMLCSRiess&\NtotalMLCSMW \\
  \\[-9pt]
  $-$ No \GALEX\ data & \NnoGALEXSALT&\NnoGALEXMLCS&\NnoGALEXMLCSRiess&\NnoGALEXMLCSMW \\
  $-$ 91T-like & \NTlikeSALT& \NTlikeMLCS& \NTlikeMLCSRiess& \NTlikeMLCSMW\\
  $-$ Highly-inclined host & \NInclinedSALT&\NInclinedMLCS&\NInclinedMLCSRiess&\NInclinedMLCSMW \\
  \hline\\[-9pt]
  Main analysis comparison sample & \NmainSALT&\NmainMLCS&\NmainMLCSRiess&\NmainMLCSMW\\
  \enddata

\tablecomments{Our MLCS2k2 $R_V=2.5$ subsample is constructed from the
intersection of the H09 $R_V=1.7$ and $R_V=3.1$ samples.  The 91T-like
SN1999gp has no \GALEX{} data, and no MLCS2k2 $R_V=1.7$ measurement
in H09.}

\end{deluxetable}

%% file: Table_2_main_table.tex
\begin{deluxetable}{lccccccccccccc}
\rotate
\centering
\setlength{\tabcolsep}{0.06in} 
\tabletypesize{\footnotesize}
\tablecolumns{14}
\tablewidth{0pt}
\tablecaption{FUV measurements of the Hubble-flow SN~Ia Sample. 
\label{tab:H09_FUV}}

\tablehead{ &  \multicolumn{3}{c} { \cHR{} (mag) } &&\multicolumn{3}{c}{\GALEX{} data}&Local&Global&Local&&& \\
Name& SALT & \multicolumn{2}{c} {MLCS2k2}& $z$& Exp. & FUV   & NUV      & $A_{\mathrm{FUV}}$& Host  & Dust  & $\log(\SFRSD)\     $ &$\pIe$& Cuts \\
& & $R_V=1.7$ &$R_V=2.5$ &         & (sec)    & (mag) & (mag)    & (mag)             & Class & Corr. &{\tiny (M$_\sun$/kpc$^2$/yr)}&(\%)&Applied }
\startdata
1990O & $  -0.14 \pm 0.19 $& $  -0.02 \pm 0.16 $& $  -0.02 \pm 0.16 $& $  0.031$& 145 & $  22.77 \pm 0.64 $& $  24.73 \pm 1.68 $& $  1.9 \pm 0.6 $& SF &  Y & $  -2.53_{-0.53}^{+0.10} $& $  22$& \\ [0.24ex]
1990af & $  -0.13 \pm 0.18 $& $  -0.28 \pm 0.19 $& $  -0.25 \pm 0.20 $& $  0.050$& 489 & $  27.35 \pm 10.11 $&   $>21.8 $& $  2.0 \pm 0.6 $& Pa &  N & $  -4.68_{-\infty}^{+0.35} $& $  100$& \\ [0.24ex]
1991U & $  -0.35 \pm 0.20 $& \nodata& \nodata& $  0.033$& 219 & $  21.43 \pm 0.36 $& $  20.63 \pm 0.10 $& $  2.2 \pm 0.6 $& $\sim$SF &  Y & $  -1.83_{-0.68}^{+0.23} $& $  6$& \\ [0.24ex]
1992J & $  -0.27 \pm 0.19 $& \nodata& \nodata& $  0.046$& 218 & $  26.61 \pm 7.93 $& $  24.22 \pm 0.80 $& $  2.0 \pm 0.6 $& Pa &  N & $  -4.47_{-\infty}^{+0.56} $& $  100$& \\ [0.24ex]
1992P & $  +0.10 \pm 0.20 $& $  +0.11 \pm 0.17 $& $  +0.14 \pm 0.18 $& $  0.026$& \nodata & no image& no image&   \nodata & $\sim$SF &  \nodata &  \nodata & \nodata & {UV}\\ [0.24ex]
1992ae & $  -0.08 \pm 0.18 $& $  -0.08 \pm 0.19 $& $  -0.08 \pm 0.20 $& $  0.075$& 558 & $  24.00 \pm 0.83 $& $  23.47 \pm 0.24 $& $  2.0 \pm 0.6 $& Pa &  N & $  -2.98_{-0.86}^{+0.29} $& $  61$& \\ [0.24ex]
1992ag & $  -0.30 \pm 0.20 $& $  -0.21 \pm 0.18 $& $  -0.23 \pm 0.19 $& $  0.026$& 184 & $  19.38 \pm 0.11 $& $  19.20 \pm 0.05 $& $  1.3 \pm 0.3 $& $\sim$SF &  Y & $  -1.58_{-0.13}^{+0.04} $& $  0$&  Incl. \\ [0.24ex]
1992bg & \nodata& $  -0.01 \pm 0.17 $& $  +0.00 \pm 0.17 $& $  0.036$& 199 & $  22.23 \pm 0.45 $& $  22.78 \pm 0.38 $& $  1.7 \pm 0.5 $& SF &  Y & $  -2.25_{-0.35}^{+0.07} $& $  10$& \\ [0.24ex]
1992bh & $  +0.12 \pm 0.18 $& $  +0.26 \pm 0.16 $& $  +0.24 \pm 0.17 $& $  0.045$& 175 & $  22.78 \pm 0.54 $& $  22.51 \pm 0.28 $& $  1.9 \pm 0.6 $& SF &  Y & $  -2.18_{-0.43}^{+0.09} $& $  11$& \\ [0.24ex]
1992bk & $  +0.15 \pm 0.28 $& $  -0.05 \pm 0.23 $& $  -0.03 \pm 0.22 $& $  0.058$& 1021 & $  26.28 \pm 1.64 $& $  25.40 \pm 0.62 $& $  2.0 \pm 0.6 $& Pa &  N & $  -4.12_{-\infty}^{+0.17} $& $  100$& \\ [0.24ex]
1992bl & $  -0.05 \pm 0.20 $& $  -0.08 \pm 0.18 $& $  -0.04 \pm 0.17 $& $  0.043$& 208 &   $>21.5 $& $  26.24 \pm 3.11 $& $  2.0 \pm 0.6 $& $\sim$Pa &  N & $<-3.8$& $  100$& \\ [0.24ex]
1992bp & $  -0.27 \pm 0.17 $& $  -0.16 \pm 0.15 $& $  -0.13 \pm 0.14 $& $  0.079$& 106 &   $>20.6 $& $  27.41 \pm 9.45 $& $  2.0 \pm 0.6 $& Pa &  N & $<-3.0$& $  66$& \\ [0.24ex]
1992br & $  +0.11 \pm 0.21 $& $  -0.63 \pm 0.22 $& $  -0.51 \pm 0.24 $& $  0.088$& \nodata & no image& no image&   \nodata & Pa &  \nodata &  \nodata & \nodata & {UV}\\ [0.24ex]
1992bs & $  +0.20 \pm 0.17 $& $  +0.23 \pm 0.18 $& $  +0.23 \pm 0.19 $& $  0.063$& 216 & $  22.45 \pm 0.40 $& $  23.13 \pm 0.32 $& $  1.6 \pm 0.5 $& SF &  Y & $  -1.87_{-0.29}^{+0.07} $& $  5$& \\ [0.24ex]
1993B & $  -0.11 \pm 0.17 $& $  +0.07 \pm 0.17 $& $  +0.10 \pm 0.17 $& $  0.071$& 192 & $  23.14 \pm 0.61 $& $  22.14 \pm 0.21 $& $  2.1 \pm 0.6 $& SF &  Y & $  -1.84_{-0.54}^{+0.10} $& $  11$& \\ [0.24ex]
1993H & \nodata& $  -0.25 \pm 0.17 $& $  -0.22 \pm 0.17 $& $  0.025$& 137 & $  23.05 \pm 0.88 $& $  22.34 \pm 0.35 $& $  2.0 \pm 0.6 $& SF &  Y & $  -2.78_{-0.83}^{+0.12} $& $  39$& \\ [0.24ex]
1993O & $  +0.02 \pm 0.17 $& $  +0.11 \pm 0.14 $& $  +0.16 \pm 0.14 $& $  0.052$& 215 &   $>21.1 $& $  25.08 \pm 1.34 $& $  2.0 \pm 0.6 $& Pa &  N & $<-3.8$& $  99$& \\ [0.24ex]
1993ac & $  +0.05 \pm 0.19 $& $  +0.06 \pm 0.19 $& $  +0.05 \pm 0.21 $& $  0.049$& 224 & $  26.19 \pm 6.78 $& $  24.36 \pm 0.78 $& $  2.0 \pm 0.6 $& Pa &  N & $  -4.24_{-\infty}^{+0.30} $& $  98$& \\ [0.24ex]
1993ag & $  -0.05 \pm 0.18 $& $  +0.23 \pm 0.16 $& $  +0.22 \pm 0.16 $& $  0.050$& \nodata & no image& no image&   \nodata & Pa &  \nodata &  \nodata & \nodata & {UV}\\ [0.24ex]
1994M & $  +0.03 \pm 0.20 $& $  -0.01 \pm 0.18 $& $  -0.02 \pm 0.18 $& $  0.024$& 109 &   $>22.5 $& $  24.97 \pm 2.68 $& $  2.0 \pm 0.6 $& $\sim$Pa &  N & $<-4.3$& $  100$& \\ [0.24ex]
1994T & $  -0.02 \pm 0.19 $& $  -0.28 \pm 0.16 $& $  -0.20 \pm 0.16 $& $  0.036$& 136 &   $>21.8 $& $  25.05 \pm 1.93 $& $  2.0 \pm 0.6 $& Pa &  N & $<-4.3$& $  100$& \\ [0.24ex]
1995ac & $  -0.32 \pm 0.17 $& $  -0.23 \pm 0.13 $& $  -0.27 \pm 0.14 $& $  0.049$& 211 & $  23.99 \pm 0.91 $& $  22.83 \pm 0.30 $& $  2.1 \pm 0.6 $& SF &  Y & $  -2.53_{-0.90}^{+0.12} $& $  26$&  Incl. \\ [0.24ex]
1996C & $  +0.20 \pm 0.20 $& $  +0.32 \pm 0.16 $& $  +0.34 \pm 0.17 $& $  0.028$& \nodata & no image& no image&   \nodata & SF &  \nodata &  \nodata & \nodata & {UV}\\ [0.24ex]
1996bl & $  -0.12 \pm 0.18 $& $  -0.00 \pm 0.15 $& $  -0.01 \pm 0.16 $& $  0.035$& 221 & $  22.15 \pm 0.35 $& $  22.18 \pm 0.21 $& $  1.7 \pm 0.5 $& SF &  Y & $  -2.24_{-0.30}^{+0.07} $& $  7$& \\ [0.24ex]
1997dg & $  +0.41 \pm 0.19 $& $  +0.38 \pm 0.16 $& $  +0.38 \pm 0.16 $& $  0.030$& 488 & $  21.63 \pm 0.22 $& $  21.96 \pm 0.13 $& $  1.2 \pm 0.4 $& $\sim$SF &  Y & $  -2.38_{-0.32}^{+0.08} $& $  5$&  Incl. \\ [0.24ex]
1998ab & $  -0.31 \pm 0.19 $& $  -0.32 \pm 0.16 $& $  -0.37 \pm 0.16 $& $  0.028$& 316 & $  20.52 \pm 0.23 $& $  20.36 \pm 0.07 $& $  1.6 \pm 0.4 $& SF &  Y & $  -1.84_{-0.69}^{+0.41} $& $  2$&  91T \\ [0.24ex]
1998dx & $  -0.10 \pm 0.18 $& $  -0.15 \pm 0.14 $& $  -0.16 \pm 0.14 $& $  0.054$& 153 &   $>21.3 $& $  26.29 \pm 3.72 $& $  2.0 \pm 0.6 $& $\sim$Pa &  N & $<-3.7$& $  97$& \\ [0.24ex]
1998eg & $  +0.07 \pm 0.21 $& $  +0.09 \pm 0.18 $& $  +0.08 \pm 0.18 $& $  0.024$& 166 & $  22.87 \pm 0.69 $& $  23.88 \pm 0.96 $& $  1.9 \pm 0.6 $& $\sim$Pa &  N & $  -3.56_{-0.44}^{+0.06} $& $  100$&  Incl. \\ [0.24ex]
1999aw \tablenotemark{b}& $  +0.07 \pm 0.18 $& \nodata& \nodata& $  0.039$& 322 & $  27.66 \pm 19.96 $&   $>22.4 $& $  2.0 \pm 0.6 $& SF &  Y & $<-3.5$& $  75$& \\ [0.24ex]
1999cc & $  -0.03 \pm 0.18 $& $  -0.05 \pm 0.15 $& $  -0.06 \pm 0.15 $& $  0.032$& 137 & $  21.93 \pm 0.40 $& $  20.92 \pm 0.14 $& $  2.2 \pm 0.6 $& SF &  Y & $  -2.03_{-0.40}^{+0.09} $& $  9$& \\ [0.24ex]
1999ef & $  +0.24 \pm 0.19 $& $  +0.33 \pm 0.16 $& $  +0.37 \pm 0.16 $& $  0.038$& 112 & $  23.98 \pm 1.31 $& $  23.14 \pm 0.52 $& $  2.0 \pm 0.6 $& SF &  Y & $  -2.78_{-\infty}^{+0.16} $& $  40$& \\ [0.24ex]
1999gp & $  +0.01 \pm 0.19 $& \nodata& \nodata& $  0.026$& \nodata & no image& no image&   \nodata & SF &  \nodata &  \nodata & \nodata &  91T {UV}\\ [0.24ex]
2000bh & $  -0.12 \pm 0.21 $& $  +0.00 \pm 0.20 $& $  +0.02 \pm 0.20 $& $  0.024$& \nodata & no image& no image&   \nodata & SF &  \nodata &  \nodata & \nodata & {UV}\\ [0.24ex]
2000ca & $  -0.16 \pm 0.20 $& $  -0.15 \pm 0.16 $& $  -0.11 \pm 0.16 $& $  0.025$& 3362 & $  20.93 \pm 0.07 $& $  20.21 \pm 0.02 $& $  2.6 \pm 0.2 $& SF &  Y & $  -1.72_{-0.40}^{+0.26} $& $  0$& \\ [0.24ex]
2000cf & $  +0.18 \pm 0.18 $& $  +0.17 \pm 0.14 $& $  +0.18 \pm 0.15 $& $  0.036$& 204 & $  21.76 \pm 0.46 $& $  21.22 \pm 0.13 $& $  2.0 \pm 0.6 $& SF &  Y & $  -1.93_{-0.76}^{+0.28} $& $  8$& \\ [0.24ex]
2001ah & $  -0.10 \pm 0.18 $& $  -0.03 \pm 0.17 $& $  +0.02 \pm 0.16 $& $  0.058$& \nodata & no image& no image&   \nodata & SF &  \nodata &  \nodata & \nodata & {UV}\\ [0.24ex]
2001az & $  +0.05 \pm 0.18 $& $  -0.04 \pm 0.15 $& $  +0.00 \pm 0.15 $& $  0.041$& \nodata & no image& $  22.62 \pm 0.28 $&   \nodata & SF &  \nodata &  \nodata & \nodata & {UV}\\ [0.24ex]
2001ba & $  +0.10 \pm 0.19 $& $  +0.08 \pm 0.15 $& $  +0.13 \pm 0.14 $& $  0.030$& 107 & $  22.38 \pm 0.63 $& $  22.53 \pm 0.42 $& $  1.9 \pm 0.6 $& SF &  Y & $  -2.39_{-0.48}^{+0.09} $& $  16$& \\ [0.24ex]
2001eh & $  -0.05 \pm 0.18 $& $  +0.12 \pm 0.13 $& $  +0.14 \pm 0.13 $& $  0.036$& \nodata & no image& no image&   \nodata & SF &  \nodata &  \nodata & \nodata & {UV}\\ [0.24ex]
2001gb & \nodata& $  +0.04 \pm 0.23 $& \nodata& $  0.027$& \nodata & no image& no image&   \nodata & SF &  \nodata &  \nodata & \nodata & {UV}\\ [0.24ex]
2001ic & \nodata& $  -0.10 \pm 0.21 $& \nodata& $  0.043$& 208 &   $>21.8 $& $  23.89 \pm 0.66 $& $  2.0 \pm 0.6 $& Pa &  N & $<-5.8$& $  100$& \\ [0.24ex]
2001ie & $  -0.02 \pm 0.20 $& $  -0.03 \pm 0.18 $& $  -0.04 \pm 0.20 $& $  0.031$& 103 &   $>22.0 $& $  25.22 \pm 2.40 $& $  2.0 \pm 0.6 $& Pa &  N & $<-3.9$& $  100$& \\ [0.24ex]
2002G & $  +0.04 \pm 0.24 $& $  -0.40 \pm 0.35 $& $  -0.41 \pm 0.38 $& $  0.035$& 173 & $  23.36 \pm 0.72 $& $  22.07 \pm 0.22 $& $  2.1 \pm 0.6 $& SF &  Y & $  -2.55_{-0.65}^{+0.11} $& $  24$& \\ [0.24ex]
2002bf & $  -0.20 \pm 0.21 $& $  +0.13 \pm 0.18 $& $  +0.11 \pm 0.19 $& $  0.025$& 112 & $  22.04 \pm 0.47 $& $  20.40 \pm 0.12 $& $  2.5 \pm 0.6 $& $\sim$SF &  Y & $  -2.18_{-0.48}^{+0.10} $& $  12$& \\ [0.24ex]
2002bz & $  +0.11 \pm 0.24 $& $  -0.01 \pm 0.18 $& \nodata& $  0.038$& \nodata & no image& no image&   \nodata & SF &  \nodata &  \nodata & \nodata & {UV}\\ [0.24ex]
2002ck & $  +0.02 \pm 0.19 $& $  +0.02 \pm 0.17 $& $  +0.03 \pm 0.17 $& $  0.030$& \nodata & no image& no image&   \nodata & $\sim$SF &  \nodata &  \nodata & \nodata & {UV}\\ [0.24ex]
2002de & $  +0.06 \pm 0.20 $& $  +0.12 \pm 0.16 $& $  +0.08 \pm 0.18 $& $  0.028$& 110 & $  20.51 \pm 0.23 $& $  19.80 \pm 0.09 $& $  2.2 \pm 0.5 $& SF &  Y & $  -1.58_{-0.29}^{+0.07} $& $  3$& \\ [0.24ex]
2002hd & $  -0.31 \pm 0.19 $& $  -0.34 \pm 0.17 $& $  -0.33 \pm 0.18 $& $  0.036$& 112 & $  23.10 \pm 0.80 $& $  21.59 \pm 0.22 $& $  2.1 \pm 0.6 $& Pa &  N & $  -3.28_{-0.51}^{+0.09} $& $  94$& \\ [0.24ex]
2002he & $  +0.08 \pm 0.21 $& $  -0.08 \pm 0.19 $& $  -0.06 \pm 0.19 $& $  0.025$& 110 & $  25.01 \pm 3.10 $& $  23.28 \pm 0.68 $& $  2.0 \pm 0.6 $& $\sim$SF &  N & $  -4.37_{-\infty}^{+0.35} $& $  100$& \\ [0.24ex]
2002hu & $  -0.11 \pm 0.18 $& $  -0.03 \pm 0.13 $& $  +0.00 \pm 0.13 $& $  0.038$& 128 & $  25.03 \pm 2.68 $& $  22.90 \pm 0.44 $& $  2.0 \pm 0.6 $& $\sim$SF &  N & $  -4.00_{-\infty}^{+0.27} $& $  100$& \\ [0.24ex]
2003D & \nodata& $  -0.26 \pm 0.19 $& $  -0.32 \pm 0.20 $& $  0.024$& 168 & $  24.11 \pm 1.60 $& $  21.76 \pm 0.22 $& $  2.1 \pm 0.6 $& Pa &  N & $  -4.10_{-\infty}^{+0.16} $& $  100$& \\ [0.24ex]
2003U & $  -0.04 \pm 0.22 $& $  -0.12 \pm 0.16 $& $  -0.08 \pm 0.16 $& $  0.028$& 170 & $  21.73 \pm 0.34 $& $  21.16 \pm 0.15 $& $  2.0 \pm 0.5 $& SF &  Y & $  -2.15_{-0.33}^{+0.08} $& $  7$& \\ [0.24ex]
2003ch & $  +0.14 \pm 0.19 $& $  +0.25 \pm 0.16 $& $  +0.24 \pm 0.16 $& $  0.030$& 204 & $  26.17 \pm 7.04 $& $  24.56 \pm 1.44 $& $  2.0 \pm 0.6 $& Pa &  N & $  -4.67_{-\infty}^{+0.45} $& $  100$& \\ [0.24ex]
2003cq & $  -0.04 \pm 0.21 $& $  +0.00 \pm 0.20 $& $  -0.06 \pm 0.24 $& $  0.034$& 81 & $  22.11 \pm 0.56 $& $  21.78 \pm 0.28 $& $  2.0 \pm 0.6 $& SF &  Y & $  -2.15_{-0.45}^{+0.09} $& $  11$& \\ [0.24ex]
2003fa & $  -0.11 \pm 0.18 $& $  +0.03 \pm 0.13 $& $  +0.04 \pm 0.12 $& $  0.039$& 128 & $  26.09 \pm 5.75 $& $  24.95 \pm 1.70 $& $  2.0 \pm 0.6 $& $\sim$SF &  N & $  -4.40_{-\infty}^{+0.53} $& $  100$& \\ [0.24ex]
2003hu & $  -0.28 \pm 0.22 $& $  -0.15 \pm 0.17 $& $  -0.13 \pm 0.18 $& $  0.075$& 294 & $  23.70 \pm 0.64 $& $  22.86 \pm 0.24 $& $  2.1 \pm 0.6 $& SF &  Y & $  -2.03_{-0.54}^{+0.10} $& $  12$&  91T \\ [0.24ex]
2003ic \tablenotemark{a}& $  -0.29 \pm 0.18 $& $  -0.27 \pm 0.16 $& $  -0.28 \pm 0.16 $& $  0.054$& 2511 & $  24.70 \pm 0.36 $& $  24.06 \pm 0.17 $& $  2.1 \pm 0.6 $& Pa &  N & $  -3.56_{-0.18}^{+0.05} $& $  100$& \\ [0.24ex]
2003it & $  +0.13 \pm 0.21 $& $  +0.04 \pm 0.19 $& $  +0.02 \pm 0.19 $& $  0.024$& 1613 & $  21.84 \pm 0.12 $& $  21.52 \pm 0.06 $& $  1.6 \pm 0.3 $& SF &  Y & $  -2.51_{-0.15}^{+0.04} $& $  4$& \\ [0.24ex]
2003iv & $  +0.18 \pm 0.20 $& $  +0.24 \pm 0.16 $& $  +0.22 \pm 0.16 $& $  0.034$& 301 & $  24.54 \pm 1.39 $& $  23.48 \pm 0.44 $& $  2.0 \pm 0.6 $& Pa &  N & $  -3.92_{-\infty}^{+0.13} $& $  100$& \\ [0.24ex]
2004L & $  +0.04 \pm 0.20 $& $  +0.11 \pm 0.17 $& $  +0.02 \pm 0.19 $& $  0.033$& 385 & $  21.00 \pm 0.23 $& $  20.65 \pm 0.07 $& $  1.8 \pm 0.4 $& SF &  Y & $  -1.80_{-0.61}^{+0.31} $& $  2$& \\ [0.24ex]
2004as & $  +0.16 \pm 0.19 $& $  +0.27 \pm 0.15 $& $  +0.27 \pm 0.16 $& $  0.032$& 106 & $  21.20 \pm 0.32 $& $  21.38 \pm 0.20 $& $  1.6 \pm 0.5 $& SF &  Y & $  -1.98_{-0.26}^{+0.07} $& $  4$& \\ [0.24ex]
2005eq & $  +0.08 \pm 0.19 $& $  +0.21 \pm 0.15 $& $  +0.26 \pm 0.15 $& $  0.028$& 1693 & $  23.74 \pm 0.35 $& $  22.73 \pm 0.12 $& $  2.3 \pm 0.6 $& $\sim$SF &  Y & $  -2.82_{-0.38}^{+0.08} $& $  38$& \\ [0.24ex]
2005eu & $  +0.01 \pm 0.19 $& $  +0.10 \pm 0.15 $& $  +0.14 \pm 0.14 $& $  0.034$& 3352 & $  22.30 \pm 0.10 $& $  22.07 \pm 0.05 $& $  1.3 \pm 0.3 $& SF &  Y & $  -2.48_{-0.13}^{+0.04} $& $  2$& \\ [0.24ex]
2005hc \tablenotemark{a}& $  +0.08 \pm 0.17 $& $  +0.11 \pm 0.14 $& $  +0.18 \pm 0.14 $& $  0.045$& 3269 & $  22.89 \pm 0.13 $& $  22.67 \pm 0.07 $& $  1.4 \pm 0.3 $& $\sim$SF &  Y & $  -2.42_{-0.15}^{+0.04} $& $  2$& \\ [0.24ex]
2005hf & $  +0.06 \pm 0.20 $& $  +0.10 \pm 0.16 $& $  +0.09 \pm 0.16 $& $  0.042$& 190 & $  26.13 \pm 4.19 $& $  23.47 \pm 0.48 $& $  2.0 \pm 0.6 $& Pa &  N & $  -4.35_{-\infty}^{+0.47} $& $  100$& \\ [0.24ex]
2005hj & $  +0.15 \pm 0.18 $& $  +0.09 \pm 0.14 $& $  +0.15 \pm 0.13 $& $  0.057$& 1675 & $  24.32 \pm 0.37 $& $  23.90 \pm 0.19 $& $  1.9 \pm 0.5 $& SF &  Y & $  -2.58_{-0.35}^{+0.08} $& $  18$& \\ [0.24ex]
2005iq & $  +0.21 \pm 0.18 $& $  +0.18 \pm 0.15 $& $  +0.22 \pm 0.15 $& $  0.033$& 112 & $  21.89 \pm 0.43 $& $  21.88 \pm 0.26 $& $  1.8 \pm 0.5 $& $\sim$SF &  Y & $  -2.15_{-0.34}^{+0.07} $& $  8$& \\ [0.24ex]
2005ir & $  +0.45 \pm 0.19 $& $  +0.24 \pm 0.14 $& $  +0.28 \pm 0.14 $& $  0.075$& 121 & $  23.13 \pm 0.73 $& $  22.22 \pm 0.27 $& $  2.1 \pm 0.6 $& SF &  Y & $  -1.81_{-0.62}^{+0.11} $& $  12$& \\ [0.24ex]
2005lz & $  +0.18 \pm 0.18 $& $  +0.18 \pm 0.15 $& $  +0.16 \pm 0.16 $& $  0.040$& \nodata & no image&   $>22.4 $&   \nodata & SF &  \nodata &  \nodata & \nodata & {UV}\\ [0.24ex]
2005mc \tablenotemark{a}& $  +0.17 \pm 0.20 $& $  +0.01 \pm 0.16 $& $  -0.03 \pm 0.16 $& $  0.026$& 1616 & $  23.49 \pm 0.28 $& $  22.06 \pm 0.08 $& $  2.9 \pm 0.5 $& $\sim$Pa &  Y & $  -2.54_{-0.31}^{+0.07} $& $  14$& \\ [0.24ex]
2005ms & $  +0.09 \pm 0.19 $& $  +0.20 \pm 0.16 $& $  +0.23 \pm 0.16 $& $  0.026$& 216 &   $>22.4 $& $  24.26 \pm 0.99 $& $  2.0 \pm 0.6 $& SF &  Y & $<-3.6$& $  98$& \\ [0.24ex]
2005na & $  -0.07 \pm 0.19 $& $  -0.14 \pm 0.16 $& $  -0.08 \pm 0.16 $& $  0.027$& \nodata & no image& no image&   \nodata & SF &  \nodata &  \nodata & \nodata & {UV}\\ [0.24ex]
2006S & $  +0.11 \pm 0.18 $& $  +0.10 \pm 0.14 $& $  +0.15 \pm 0.15 $& $  0.033$& 96 & $  21.53 \pm 0.39 $& $  21.57 \pm 0.23 $& $  1.8 \pm 0.5 $& SF &  Y & $  -2.03_{-0.33}^{+0.08} $& $  6$& \\ [0.24ex]
2006ac & $  -0.06 \pm 0.20 $& $  +0.01 \pm 0.17 $& $  +0.03 \pm 0.17 $& $  0.024$& 213 & $  20.19 \pm 0.14 $& $  19.99 \pm 0.07 $& $  1.4 \pm 0.4 $& SF &  Y & $  -1.92_{-0.16}^{+0.05} $& $  1$& \\ [0.24ex]
2006ak & $  +0.00 \pm 0.22 $& $  -0.04 \pm 0.19 $& $  +0.01 \pm 0.18 $& $  0.039$& 198 & $  22.03 \pm 0.46 $& $  21.93 \pm 0.19 $& $  1.8 \pm 0.5 $& $\sim$SF &  Y & $  -2.04_{-0.61}^{+0.17} $& $  8$&  Incl. \\ [0.24ex]
2006al & $  +0.02 \pm 0.18 $& $  +0.19 \pm 0.14 $& $  +0.21 \pm 0.14 $& $  0.069$& 1575 & $  26.37 \pm 1.20 $&   $>21.7 $& $  2.0 \pm 0.6 $& Pa &  N & $  -4.00_{-\infty}^{+0.13} $& $  100$& \\ [0.24ex]
2006an \tablenotemark{b}& $  -0.05 \pm 0.17 $& $  +0.18 \pm 0.14 $& $  +0.21 \pm 0.13 $& $  0.065$& 93 &   $>21.1 $& $  24.05 \pm 0.83 $& $  2.0 \pm 0.6 $& SF &  Y & $<-2.2$& $  23$& \\ [0.24ex]
2006az & $  -0.06 \pm 0.18 $& $  -0.08 \pm 0.14 $& $  -0.05 \pm 0.14 $& $  0.032$& 187 & $  23.19 \pm 0.64 $& $  22.08 \pm 0.22 $& $  2.1 \pm 0.6 $& Pa &  N & $  -3.43_{-0.39}^{+0.07} $& $  100$& \\ [0.24ex]
2006bd & \nodata& $  +0.14 \pm 0.17 $& $  +0.16 \pm 0.19 $& $  0.026$& 1607 & $  25.73 \pm 1.16 $& $  23.60 \pm 0.19 $& $  2.1 \pm 0.6 $& Pa &  N & $  -4.50_{-\infty}^{+0.12} $& $  100$& \\ [0.24ex]
2006bt & $  -0.01 \pm 0.18 $& $  +0.05 \pm 0.14 $& $  -0.07 \pm 0.15 $& $  0.033$& 205 &   $>22.0 $&   $>22.7 $& $  2.0 \pm 0.6 $& $\sim$SF &  N & $<-4.2$& $  100$& \\ [0.24ex]
2006bu & $  +0.07 \pm 0.23 $& $  -0.10 \pm 0.14 $& $  -0.06 \pm 0.13 $& $  0.084$& 1675 &   $>20.4 $& $  26.17 \pm 0.65 $& $  2.0 \pm 0.6 $& SF &  Y & $<-3.7$& $  99$& \\ [0.24ex]
2006bw & $  -0.04 \pm 0.21 $& $  -0.15 \pm 0.18 $& $  -0.14 \pm 0.19 $& $  0.031$& 1696 &   $>22.1 $& $  25.62 \pm 0.95 $& $  2.0 \pm 0.6 $& Pa &  N & $<-4.9$& $  100$& \\ [0.24ex]
2006bz & \nodata& $  -0.20 \pm 0.16 $& $  -0.24 \pm 0.18 $& $  0.028$& 25656 & $  25.45 \pm 0.27 $& $  23.95 \pm 0.06 $& $  3.1 \pm 0.4 $& Pa &  N & $  -4.45_{-0.27}^{+0.10} $& $  100$& \\ [0.24ex]
2006cc & $  +0.27 \pm 0.18 $& $  +0.25 \pm 0.14 $& $  +0.01 \pm 0.15 $& $  0.033$& 904 & $  22.97 \pm 0.26 $& $  22.26 \pm 0.11 $& $  2.2 \pm 0.5 $& $\sim$SF &  Y & $  -2.45_{-0.31}^{+0.07} $& $  11$&  Incl. \\ [0.24ex]
2006cf & $  -0.03 \pm 0.20 $& $  -0.01 \pm 0.15 $& $  +0.04 \pm 0.15 $& $  0.042$& 108 & $  21.48 \pm 0.36 $& $  21.29 \pm 0.19 $& $  1.8 \pm 0.5 $& SF &  Y & $  -1.77_{-0.31}^{+0.07} $& $  5$& \\ [0.24ex]
2006cg & $  -0.50 \pm 0.26 $& $  -0.55 \pm 0.18 $& $  -0.51 \pm 0.20 $& $  0.029$& 1391 & $  24.97 \pm 0.67 $& $  23.47 \pm 0.19 $& $  2.2 \pm 0.6 $& Pa &  N & $  -4.22_{-0.41}^{+0.07} $& $  100$& \\ [0.24ex]
2006cj & $  +0.29 \pm 0.18 $& $  +0.20 \pm 0.13 $& $  +0.23 \pm 0.13 $& $  0.068$& 25656 & $  23.85 \pm 0.08 $& $  23.39 \pm 0.03 $& $  1.8 \pm 0.3 $& $\sim$SF &  Y & $  -2.29_{-0.26}^{+0.10} $& $  0$& \\ [0.24ex]
2006cq & $  +0.05 \pm 0.18 $& $  +0.18 \pm 0.16 $& $  +0.21 \pm 0.17 $& $  0.049$& 1660 & $  23.77 \pm 0.27 $& $  23.01 \pm 0.11 $& $  2.2 \pm 0.5 $& SF &  Y & $  -2.39_{-0.32}^{+0.08} $& $  10$& \\ [0.24ex]
2006cs & \nodata& $  -0.00 \pm 0.18 $& $  -0.03 \pm 0.21 $& $  0.024$& 106 & $  23.57 \pm 1.10 $& $  23.51 \pm 0.75 $& $  2.0 \pm 0.6 $& Pa &  N & $  -3.79_{-0.70}^{+0.13} $& $  100$& \\ [0.24ex]
2006en & $  +0.10 \pm 0.19 $& $  +0.12 \pm 0.17 $& $  +0.12 \pm 0.19 $& $  0.031$& 265 & $  21.04 \pm 0.19 $& $  20.08 \pm 0.07 $& $  2.7 \pm 0.5 $& SF &  Y & $  -1.51_{-0.27}^{+0.07} $& $  2$& \\ [0.24ex]
2006gj & $  +0.27 \pm 0.20 $& $  +0.09 \pm 0.17 $& $  -0.12 \pm 0.17 $& $  0.028$& 224 & $  24.21 \pm 1.44 $& $  23.84 \pm 0.73 $& $  2.0 \pm 0.6 $& $\sim$Pa &  N & $  -3.96_{-\infty}^{+0.16} $& $  100$&  Incl. \\ [0.24ex]
2006gr & $  +0.09 \pm 0.18 $& $  +0.19 \pm 0.14 $& $  +0.19 \pm 0.15 $& $  0.034$& \nodata & no image& no image&   \nodata & SF &  \nodata &  \nodata & \nodata & {UV}\\ [0.24ex]
2006gt & \nodata& $  +0.14 \pm 0.17 $& $  +0.18 \pm 0.18 $& $  0.044$& 102 &   $>21.5 $& $  23.85 \pm 0.80 $& $  2.0 \pm 0.6 $& Pa &  N & $<-3.5$& $  95$& \\ [0.24ex]
2006mo & $  +0.17 \pm 0.20 $& $  -0.01 \pm 0.16 $& $  +0.02 \pm 0.17 $& $  0.036$& 1704 & $  24.44 \pm 0.49 $& $  23.21 \pm 0.14 $& $  2.3 \pm 0.6 $& Pa &  N & $  -3.81_{-0.27}^{+0.04} $& $  100$& \\ [0.24ex]
2006nz & $  +0.21 \pm 0.22 $& $  -0.20 \pm 0.18 $& $  -0.18 \pm 0.18 $& $  0.037$& 3007 & $  24.41 \pm 0.32 $& $  23.06 \pm 0.09 $& $  2.7 \pm 0.6 $& Pa &  N & $  -3.77_{-0.17}^{+0.04} $& $  100$& \\ [0.24ex]
2006oa & $  -0.00 \pm 0.17 $& $  +0.01 \pm 0.14 $& $  +0.06 \pm 0.14 $& $  0.059$& 3121 & $  23.64 \pm 0.19 $& $  23.63 \pm 0.12 $& $  1.4 \pm 0.4 $& SF &  Y & $  -2.50_{-0.20}^{+0.05} $& $  7$& \\ [0.24ex]
2006ob & $  +0.02 \pm 0.17 $& $  -0.11 \pm 0.14 $& $  -0.10 \pm 0.13 $& $  0.058$& 3279 & $  25.44 \pm 0.48 $& $  24.49 \pm 0.19 $& $  2.1 \pm 0.6 $& SF &  Y & $  -2.93_{-0.45}^{+0.09} $& $  53$& \\ [0.24ex]
2006on & $  -0.02 \pm 0.20 $& $  -0.03 \pm 0.18 $& $  -0.09 \pm 0.19 $& $  0.069$& 3934 &   $>20.8 $& $  25.36 \pm 0.32 $& $  2.0 \pm 0.6 $& Pa &  N & $<-4.7$& $  100$& \\ [0.24ex]
2006os & $  -0.11 \pm 0.19 $& $  -0.11 \pm 0.18 $& $  -0.36 \pm 0.19 $& $  0.032$& 110 & $  23.31 \pm 1.05 $& $  22.41 \pm 0.38 $& $  2.0 \pm 0.6 $& $\sim$SF &  N & $  -3.47_{-1.09}^{+0.11} $& $  99$& \\ [0.24ex]
2006qo & $  -0.03 \pm 0.19 $& $  -0.04 \pm 0.16 $& $  -0.14 \pm 0.16 $& $  0.030$& \nodata & no image& no image&   \nodata & SF &  \nodata &  \nodata & \nodata & {UV}\\ [0.24ex]
2006te & $  +0.10 \pm 0.18 $& $  +0.11 \pm 0.15 $& $  +0.15 \pm 0.15 $& $  0.032$& 196 & $  21.96 \pm 0.34 $& $  20.88 \pm 0.12 $& $  2.4 \pm 0.6 $& SF &  Y & $  -1.97_{-0.39}^{+0.08} $& $  7$& \\ [0.24ex]
2007F & $  +0.10 \pm 0.20 $& $  +0.11 \pm 0.16 $& $  +0.16 \pm 0.16 $& $  0.024$& 205 & $  20.35 \pm 0.21 $& $  20.29 \pm 0.09 $& $  1.5 \pm 0.4 $& SF &  Y & $  -1.94_{-0.48}^{+0.22} $& $  2$& \\ [0.24ex]
2007H & $  +0.14 \pm 0.27 $& \nodata& \nodata& $  0.044$& \nodata & no image& no image&   \nodata & $\sim$SF &  \nodata &  \nodata & \nodata & {UV}\\ [0.24ex]
2007O & $  -0.08 \pm 0.18 $& $  -0.03 \pm 0.15 $& $  +0.04 \pm 0.15 $& $  0.036$& 183 & $  20.78 \pm 0.20 $& $  20.27 \pm 0.09 $& $  2.0 \pm 0.5 $& SF &  Y & $  -1.57_{-0.24}^{+0.06} $& $  2$& \\ [0.24ex]
2007R & $  +0.22 \pm 0.20 $& $  +0.07 \pm 0.16 $& $  +0.12 \pm 0.16 $& $  0.031$& 211 & $  21.01 \pm 0.21 $& $  20.26 \pm 0.08 $& $  2.3 \pm 0.5 $& $\sim$SF &  Y & $  -1.66_{-0.28}^{+0.07} $& $  3$& \\ [0.24ex]
2007ae & $  -0.20 \pm 0.18 $& $  -0.16 \pm 0.15 $& $  -0.13 \pm 0.14 $& $  0.064$& 185 & $  23.40 \pm 0.70 $& $  23.96 \pm 0.56 $& $  1.9 \pm 0.6 $& $\sim$SF &  N & $  -2.88_{-0.44}^{+0.07} $& $  47$& \\ [0.24ex]
2007ai & $  +0.03 \pm 0.20 $& $  +0.11 \pm 0.18 $& $  +0.10 \pm 0.19 $& $  0.032$& \nodata & no image& no image&   \nodata & SF &  \nodata &  \nodata & \nodata & {UV}\\ [0.24ex]
2007ar & \nodata& $  -0.53 \pm 0.18 $& \nodata& $  0.053$& \nodata & no image& no image&   \nodata & SF &  \nodata &  \nodata & \nodata & {UV}\\ [0.24ex]
2007ba & \nodata& $  -0.52 \pm 0.15 $& $  -0.46 \pm 0.15 $& $  0.039$& 2857 & $  23.70 \pm 0.23 $& $  23.39 \pm 0.12 $& $  1.8 \pm 0.5 $& Pa &  N & $  -3.64_{-0.11}^{+0.03} $& $  100$& \\ [0.24ex]
2007bd & $  -0.09 \pm 0.19 $& $  -0.16 \pm 0.17 $& $  -0.09 \pm 0.16 $& $  0.032$& 109 & $  21.37 \pm 0.35 $& $  21.09 \pm 0.17 $& $  1.8 \pm 0.5 $& $\sim$SF &  Y & $  -1.95_{-0.31}^{+0.07} $& $  5$& \\ [0.24ex]
2007cg & $  -0.18 \pm 0.21 $& $  -0.30 \pm 0.20 $& \nodata& $  0.034$& 111 & $  21.76 \pm 0.43 $& $  21.31 \pm 0.19 $& $  2.0 \pm 0.6 $& $\sim$SF &  Y & $  -2.01_{-0.37}^{+0.09} $& $  8$& \\ [0.24ex]
2007co & $  -0.03 \pm 0.19 $& $  -0.08 \pm 0.16 $& $  -0.11 \pm 0.16 $& $  0.027$& \nodata & no image& no image&   \nodata & SF &  \nodata &  \nodata & \nodata & {UV}\\ [0.24ex]
2007cq & $  -0.25 \pm 0.20 $& $  -0.25 \pm 0.17 $& $  -0.20 \pm 0.18 $& $  0.025$& 435 & $  21.51 \pm 0.19 $& $  21.05 \pm 0.09 $& $  1.9 \pm 0.4 $& $\sim$Pa &  Y & $  -2.22_{-0.23}^{+0.06} $& $  4$& \\ [0.24ex]
2008af & $  -0.12 \pm 0.19 $& $  -0.03 \pm 0.17 $& $  +0.03 \pm 0.17 $& $  0.034$& 73 & $  25.69 \pm 5.21 $& $  26.38 \pm 7.58 $& $  2.0 \pm 0.6 $& Pa &  N & $  -4.37_{-\infty}^{+0.57} $& $  100$& \\ [0.24ex]
2008bf & $  -0.35 \pm 0.20 $& $  -0.25 \pm 0.16 $& $  -0.20 \pm 0.16 $& $  0.026$& 104 &   $>22.4 $& $  23.53 \pm 0.80 $& $  2.0 \pm 0.6 $& Pa &  N & $<-4.3$& $  100$& \\ [0.24ex]
\enddata
\tablecomments{The asymetric errors on $\log(\SFRSD)$ indicate the 16\% and 64\% boundaries of the \SFRSD\ cumulative probability distribution functions; a lower boundary of zero is indicated by $-\infty$ when in $\log$ space. For cases with no \GALEX\ FUV counts we only indicate the upper boundary. The uncertainties on the FUV and NUV magnitudes have been symmetrized for readability but are not directly used. The \emph{Cuts Applied} column indicates reasons why a SN~Ia was removed from the main analysis; UV stands for no \GALEX{} data, Incl. for inclined host and 91T for 91T-like SN (see Appendix~\ref{sec:R13-cuts}). The \emph{Global Host Type} column gives the global star formation classification, as defined in Section~\ref{sec:local-dust-correction}. The \emph{Local $A_{\mathrm{FUV}}$} column lists the FUV extinction, regardless of whether it is applied (as indicated by the \emph{Local Dust Corr} column). The \emph{Exp.} column indicates the \GALEX\ exposure time in seconds.}
\tablenotetext{a}{See Appendix~~\ref{sec:bright_core}}
\tablenotetext{b}{See Appendix~~\ref{sec:dilute_signal}}
\end{deluxetable}

%% file: Table_3_SF_bias_summary.tex
\begin{deluxetable}{lcccccc}
  \centering
  \tabletypesize{\footnotesize}
  \tablecolumns{7}
  \tablewidth{0pt}
  \tablecaption{The SF bias and Environmental variations in the H09/\GALEX\ sample.
    \label{tab:SFbias_Summary}}

  \tablehead{ 
    &  \multicolumn{3}{c}{\textsc{Star-Formation Bias}} &\multicolumn{3}{c}{\textsc{SN
      Hubble~Residual dispersion\tablenotemark{a}}}\\[2pt]
    Light-curve fitter & Fraction of & \DcHR{} & Bias & SNe~Ia & SNe~\sS{} & SNe~\sN{}\\
    & SNe\ \sN{} ($\psi$) & (mag) &Signifiance& (mag) & (mag) & (mag)
  }
  \startdata
  SALT2 & \HgalexglobalHFSALT & \SFHickenNomagSALT & \SFHickenLEVELSALT &
  \IaDispersionNoPecWithErrorNomagSALT&\IaaDispersionNoPecWithErrorNomagSALT&\IaeDispersionNoPecWithErrorNomagSALT\\[4pt]
  MLCS2k2 $R_V=1.7$ & \HgalexglobalHFMLCS & \SFHickenNomagMLCS&
  \SFHickenLEVELMLCS &
  \IaDispersionNoPecWithErrorNomagMLCS&\IaaDispersionNoPecWithErrorNomagMLCS&\IaeDispersionNoPecWithErrorNomagMLCS\\
  MLCS2k2 $R_V=2.5$& \HgalexglobalHFMLCSRiess &
  \SFHickenNomagMLCSRiess & \SFHickenLEVELMLCSRiess  &
  \IaDispersionNoPecWithErrorNomagMLCSRiess&\IaaDispersionNoPecWithErrorNomagMLCSRiess&\IaeDispersionNoPecWithErrorNomagMLCSRiess\\
  MLCS2k2 $R_V=3.1$ & \HgalexglobalHFMLCSMW &
  \SFHickenNomagMLCSMW&\SFHickenLEVELMLCSMW  &
  \IaDispersionNoPecWithErrorNomagMLCSMW&\IaaDispersionNoPecWithErrorNomagMLCSMW&\IaeDispersionNoPecWithErrorNomagMLCSMW\\
  \enddata
\tablenotetext{a}{\ Weighted RMS with noise due to small-scale galaxy
  peculiar velocities ($300\,\mathrm{km\ s^{-1}}$) removed.}
  
\end{deluxetable}

%% file: Table_4_Cepheid_table.tex
\begin{deluxetable}{lccccccc}
\centering
\tabletypesize{\footnotesize}
\tablecolumns{8}
\tablewidth{0pt}
\tablecaption{Local UV Environments of the Cepheid--SNe~Ia sample. 
\label{tab:SHOES}}
\tablehead{Name & FUV    & NUV    & $A_{FUV}$& Host &  Dust  & $\log(\SFRSD)\ $     & $\pIe$  \\
                & (mag)  & (mag)  & (mag)             & Type &  Corr. & (M$_\sun$/kpc$^2$/yr)& (percent)}
\startdata
SN1981B & $  17.34 \pm 0.02 $& $  16.91 \pm 0.01 $& $  1.7 \pm 0.1 $&   SF &   Y & $  -2.34 \pm 0.02 $& $  0$\\ [0.1ex]
SN1990N & $  18.76 \pm 0.10 $& $  18.46 \pm 0.04 $& $  1.5 \pm 0.3 $&   SF &   Y & $  -2.67 \pm 0.13 $& $  9$\\ [0.1ex]
SN1994ae & $  18.33 \pm 0.08 $& $  17.77 \pm 0.04 $& $  2.1 \pm 0.3 $&   SF &   Y & $  -2.09 \pm 0.11 $& $  0$\\ [0.1ex]
SN1995al & $  16.49 \pm 0.03 $& $  15.87 \pm 0.01 $& $  2.3 \pm 0.1 $&   SF &   Y & $  -1.44 \pm 0.03 $& $  0$\\ [0.1ex]
SN1998aq & no image                & $  16.29 \pm 0.00 $&   \nodata& SF &   \nodata & $>-2.4$\tablenotemark{a}& $  0$\\ [0.1ex]
SN2002fk & $  16.46 \pm 0.01 $& $  15.87 \pm 0.00 $& $  2.2 \pm 0.0 $&   SF &   Y & $  -1.09 \pm 0.01 $& $  0$\\ [0.1ex]
SN2007af & $  17.67 \pm 0.02 $& $  17.27 \pm 0.01 $& $  1.6 \pm 0.1 $&   SF &   Y & $  -2.18 \pm 0.03 $& $  0$\\ [0.1ex]
SN2007sr \tablenotemark{b}& $  22.08 \pm 0.57 $& $  20.65 \pm 0.13 $& $  2.3 \pm 0.6 $&   SF &   Y & $  -3.68 \pm 0.33 $& $  50$\tablenotemark{c}\\ [0.1ex]
\hline
\hline\\[-9pt]
SN2001el & $  16.85 \pm 0.03 $& $  16.41 \pm 0.01 $& $  1.7 \pm 0.1 $&   SF &   Y & $  -1.98 \pm 0.04 $& $  0$\\ [0.1ex]
SN2003du & $  18.66 \pm 0.10 $& $  18.44 \pm 0.04 $& $  1.3 \pm 0.3 $&   SF &   Y & $  -2.29 \pm 0.11 $& $  0$\\ [0.1ex]
SN2005cf \tablenotemark{b}& $  22.29 \pm 0.25 $& $  21.16 \pm 0.06 $& $  2.8 \pm 0.5 $&   SF &   Y & $  -3.34 \pm 0.23 $& $  100$\\ [0.1ex]
SN2011fe & $  14.95 \pm 0.01 $& $  14.62 \pm 0.01 $& $  1.3 \pm 0.1 $&   SF &   Y & $  -2.26 \pm 0.02 $& $  0$\\ [0.1ex]
SN2012cg & $  16.67 \pm 0.01 $& $  15.91 \pm 0.00 $& $  2.7 \pm 0.0 $&   SF &   Y & $  -1.62 \pm 0.01 $& $  0$\\ [0.1ex]
SN2012fr & $  17.54 \pm 0.01 $& $  17.02 \pm 0.00 $& $  2.0 \pm 0.0 $&   SF &   Y & $  -2.15 \pm 0.02 $& $  0$\\ [0.1ex]
SN2012ht & $  20.42 \pm 0.04 $& $  19.68 \pm 0.01 $& $  2.7 \pm 0.1 $&   SF &   Y & $  -2.20 \pm 0.05 $& $  0$\\ [0.1ex]
SN2013dy & $  15.64 \pm 0.00 $& $  15.31 \pm 0.00 $& $  1.4 \pm 0.0 $&   SF &   Y & $  -1.87 \pm 0.00 $& $  0$\\ [0.1ex]
\enddata
\tablenotetext{a}{\ Based on NUV magnitude, assuming $\mathrm{FUV-NUV}=1$ and $A_{FUV}=0$.}
\tablenotetext{b}{\ Tails of interacting galaxies, \pIe\ could be problematic.}
\tablenotetext{c}{\ Revised, see Section~\ref{sec:Cepheid_psiCp}.}
\end{deluxetable}

%% file: Table_5_H0_environmental_bias.tex
\begin{deluxetable}{lcccc}
  \centering
  \tabletypesize{\small}
  \tablecolumns{5}
  \tablewidth{0pt}
  \tablecaption{Effect of SN~Ia environmental bias on \Hnot{}.
    \label{tab:environmental_H0_bias}}
  \tablehead{Component& SALT2 & \multicolumn{3}{c}{MLCS2k2}\\
    & &$R_V=1.7$&$R_V=2.5$&$R_V=3.1$
}
  \startdata 
  Host-mass correction\tablenotemark{a} &\multicolumn{4}{c}{$+0.75\%$}\\
  \textbf{SF~bias correction} & $-$\HnotbiasSALT &  $-$\HnotbiasMLCS &  $-$\HnotbiasMLCSRiess &  $-$\HnotbiasMLCSMW\\
  \\[-9pt]
  Net bias & $-$\EffBiasPercentRiessSALTMassRmv  &
  $-$\EffBiasPercentRiessMLCSMassRmv
& $-$\EffBiasPercentRiessMLCSRiessMassRmv & $-$\EffBiasPercentRiessMLCSMWMassRmv \\ 
  \enddata
  \tablenotetext{a}{\ Removal of the host-mass corrections applied by
  \shoes{} (see their Section~3.1) since the SF bias already accounts
  for the host-mass bias. We assume that the bias used for MLCS2K2 was
  also used for the SALT analysis.}
\end{deluxetable}

%% file: Table_6_H0_corr_table.tex
\begin{deluxetable}{lcc}
  \centering
  \tabletypesize{\small}
  \tablecolumns{3}
  \tablewidth{0pt}
  \tablecaption{Direct measurement of the Hubble constant (in
    $\mathrm{km\ s^{-1}\ Mpc^{-1}}$). 
  \emph{Top:} Using Cepheid distances calibrated to
  the megamaser NGC~4258, Milky Way parallaxes, 
  and LMC distance. \emph{Bottom:}  Using Cepheid distances
  calibrated soley with the distance to the NGC~4258 megamaser.  
    \label{tab:H0_corr}}
  \tablehead{ & \citealt{riess_3_2011} & \citealt{Efstathiou_H0_2014}}
  \startdata
  \\[-15pt]
  \multicolumn{3}{c}{\textsc{anchors: ngc4258,  milky way parallaxes, and lmc distance}}\\[5pt]
  \hline \\[-9pt] 
  Starting \Hnot{}                & $72.7 \pm 2.4$\tablenotemark{a} & $72.5 \pm 2.5$ \\
  No mass~bias correction &  $+$\MassBiasAmplRiessMLCSRiessMassRmvNoUnit   &     $+$\MassBiasAmplEftaTMLCSRiessMassRmvNoUnit \\
  SF~bias correction\tablenotemark{c}&  $-$\HnotBiasAmplRiessMLCSRiessMassRmvNoUnit   &  $-$\HnotBiasAmplEftaTMLCSRiessMassRmvNoUnit \\
  \textbf{Revised \Hnot{}}    & \HnotCorrRiessMLCSRiessMassRmvNoUnit     &      \HnotCorrEftaTMLCSRiessMassRmvNoUnit \\[5pt]

  New CMB tension\tablenotemark{d} & \CMBTensionRiessMLCSRiessMassRmv &  \CMBTensionEftaTMLCSRiessMassRmv\\[2pt]

  \hline \\[-9pt]
  \multicolumn{3}{c}{\textsc{anchor: ngc4258}}\\[4pt]
  \hline \\[-9pt]

  Starting \Hnot{} & $72.0 \pm 3.0$\tablenotemark{b} & $70.6 \pm 3.3$ \\
  No mass~bias correction&  $+$\MassBiasAmplHumphMLCSRiessMassRmvNoUnit   &     $+$\MassBiasAmplEftaOMLCSRiessMassRmvNoUnit \\
  SF~bias correction\tablenotemark{c}&  $-$\HnotBiasAmplHumphMLCSRiessMassRmvNoUnit &      $-$\HnotBiasAmplEftaOMLCSRiessMassRmvNoUnit \\
  
  \textbf{Revised \Hnot{}} & \HnotCorrHumphMLCSRiessMassRmvNoUnit & \HnotCorrEftaOMLCSRiessMassRmvNoUnit  \\[5pt]

  New CMB tension\tablenotemark{d} & \CMBTensionHumphMLCSRiessMassRmv &  \CMBTensionEftaOMLCSRiessMassRmv\\
  
  \enddata
  \tablenotetext{a}{\ After including the recalibration of NGC~4258 
    \citep[$-1.1\ \mathrm{km\ s^{-1}\
      Mpc^{-1}}$;][]{humphreys_correction_2013} and using the prescription given in
    Section~3.1 of \shoes.} 
  \tablenotetext{b}{\ From \cite{humphreys_correction_2013}.} 
  \tablenotetext{c}{\ Add an extra $-0.3\ \mathrm{km\ s^{-1}\
      Mpc^{-1}}$ if SN2007sr is considered to be \sS{}.}
  \tablenotetext{d}{\ Tensions with Planck \citep[$67.3 \pm 1.2\
    \mathrm{km\ s^{-1}\
      Mpc^{-1}}$;][]{planck_collaboration_planck_2013}, WMAP 9-year
    \citep[$69.32 \pm 0.80\ \mathrm{km\ s^{-1}\
      Mpc^{-1}}$;][]{bennett_WMAP9, hinshaw_WMAP9}, and revised Planck
    \citep[$68.0 \pm 1.1\ \mathrm{km\ s^{-1}\
      Mpc^{-1}}$;][]{spergel_New_Planck}, respectively.}
\end{deluxetable}

%% file: Rigault_SF_bias_H0_paper.bbl
\begin{thebibliography}{}

\bibitem[{Aldering {et~al.}(2002)}]{aldering_overview_2002}
Aldering, G., Adam, G., Antilogus, P., {et~al.} 2002, SPIE, 4836, 61


\bibitem[Baldwin et al.(1981)]{bpt} Baldwin, J.~A., 
Phillips, M.~M., \& Terlevich, R.\ 1981, \pasp, 93, 5 

\bibitem[Barbon et al.(1999)]{asiago_1999} Barbon, R., Buond{\'{\i}},
  V., Cappellaro, E., \& Turatto, M.\ 1999, \aaps, 139, 531

\bibitem[Bell
\& Kennicutt(2001)]{bell_kennicutt_2001} Bell, E.~F., \& Kennicutt,
R.~C., Jr.\ 2001, \apj, 548, 681

\bibitem[Barnes et al.(2011)]{barnes_uvdisk_2011} Barnes, K.~L., van Zee,
   L., \& Skillman, E.~D.\ 2011, \apj, 743, 137

\bibitem[Bennett et al.(2013)]{bennett_WMAP9} Bennett, C.~L., Larson,
D., Weiland, J.~L., et al.\ 2013, \apjs, 208, 20

\bibitem[Bennett et al.(2014)]{bennett_concordance} Bennett, C.~L., Larson, 
D., Weiland, J.~L., \& Hinshaw, G.\ 2014, \apj, 794, 135 

\bibitem[Bird et al.(2012)]{2012MNRAS.420..913B} Bird, J.~C., Kazantzidis, 
S., \& Weinberg, D.~H.\ 2012, \mnras, 420, 913

\bibitem[Blondin et al.(2012)]{Blondin_specto_2012} Blondin, S., Matheson,
T., Kirshner, R.~P., et al.\ 2012, \aj, 143, 126

\bibitem[Boquien et al.(2011)]{boquien_threshold_2011} Boquien, M., Calzetti, 
D., Combes, F., et al.\ 2011, \aj, 142, 111 

\bibitem[Brunetti et al.(2011)]{2011A&A...534A..75B} Brunetti, M.,
Chiappini, C., \& Pfenniger, D.\ 2011, \aap, 534, A75

\bibitem[Calzetti(2013)]{calzetti_sfr_2013} Calzetti, D.\ 2013, Secular
Evolution of Galaxies, 419

\bibitem[Cardelli et al.(1989)]{cardelli_dust_law} Cardelli, J.~A., 
Clayton, G.~C., \& Mathis, J.~S.\ 1989, \apj, 345, 245 

\bibitem[{Charlot \& Fall(2000)}]{charlot_simple_2000}
Charlot, S. \& Fall, S.~M. 2000, \apj, 539, 718

\bibitem[Childress et al.(2011)]{childress_sn2007if_2011} Childress, M., 
Aldering, G., Aragon, C., et al.\ 2011, \apj, 733, 3 

\bibitem[{Childress {et~al.}(2013a)}]{childress_sys_2013}
Childress, M., Aldering, G., Antilogus, P., {et~al.} 2013a, \apj,
  770, 108

\bibitem[Childress et al.(2013b)]{childress_data_2013} Childress, M., 
Aldering, G., Antilogus, P., et al.\ 2013b, \apj, 770, 107 

\bibitem[Chilingarian \& Zolotukhin(2012)]{2012MNRAS.419.1727C}
Chilingarian, I.~V., \& Zolotukhin, I.~Y.\ 2012, \mnras, 419, 1727

\bibitem[Conley et al.(2011)]{conley_cosmo_2011} Conley, A., Guy, J., 
Sullivan, M., et al.\ 2011, \apjs, 192, 1

\bibitem[Conroy et al.(2010)]{conroy2010} Conroy, C., 
Schiminovich, D., \& Blanton, M.~R.\ 2010, \apj, 718, 184 


\bibitem[de Vaucouleurs et al.(1991)]{rc3} de Vaucouleurs G., 
de Vaucouleurs A., Corwin H.~G., Buta R.~J., Paturel G., Fouque P., 
``Third Reference Catalogue of Bright Galaxies (RC3)''
Springer-Verlag: New York, (1991)

\bibitem[{{D'Andrea} {et~al.}(2011)}]{dandrea_spectroscopic_2011}
{D'Andrea}, C.~B., Gupta, R.~R., Sako, M., {et~al.} 2011, ApJ, 743,
172

\bibitem[de Zeeuw et al.(1999)]{dezeeuw_1999} de Zeeuw, P.~T., 
Hoogerwerf, R., de Bruijne, J.~H.~J., Brown, A.~G.~A., 
\& Blaauw, A.\ 1999, \aj, 117, 354 

\bibitem[Di Matteo et al.(2013)]{2013A&A...553A.102D} Di Matteo,
P., Haywood, M., Combes, F., Semelin, B., \& Snaith, O.~N.\ 2013,
\aap, 553, A102

\bibitem[Efstathiou(2014)]{Efstathiou_H0_2014} Efstathiou, G.\ 2014, 
\mnras, 440, 1138 
\bibitem[Fall et al.(2005)]{fall_cluster_ages_2005} Fall, S.~M., Chandar, R., 
\& Whitmore, B.~C.\ 2005, \apjl, 631, L133 

\bibitem[F{\"o}rster \& Schawinski(2008)]{forster_SFinETG_2008} 
F{\"o}rster, F., \& Schawinski, K.\ 2008, \mnras, 388, L74 

\bibitem[{Freedman {et~al.}(2001)}]{freedman_final_2001}
Freedman, W.~L. and Madore, B.~F. and Gibson, B.~K., {et~al.} 2001,
\apj, 553, 47

\bibitem[Freedman et al.(2012)]{freedman_2012} Freedman, W.~L.,
Madore, B.~F., Scowcroft, V., et al.\ 2012, \apj, 758, 24


\bibitem[Gil de Paz et al.(2007)]{gildepaz_galex_nga_2007} Gil de Paz, A., 
Boissier, S., Madore, B.~F., et al.\ 2007, \apjs, 173, 185 

\bibitem[{Gupta {et~al.}(2011)}]{gupta_improved_2011}
Gupta, R.~R., {D'Andrea}, C.~B., Sako, M., {et~al.} 2011, ApJ, 740, 92

\bibitem[{Guy {et~al.}(2007)}]{guy-salt2-2007}
Guy, J., Astier, P., Baumont, S., {et~al.} 2007, \aap, 466, 11

\bibitem[Hamuy et al.(1996)]{hamuy1996} Hamuy, M., Phillips, 
M.~M., Suntzeff, N.~B., et al.\ 1996, \aj, 112, 2408 

\bibitem[Han et al.(2007)]{han2007_FUV_ssp} Han, Z., Podsiadlowski, P.,
\& Lynas-Gray, A.~E.\ 2007, \mnras, 380, 1098

\bibitem[Hibbard et al.(2005)]{hibbard_blue_tail_2005} Hibbard, J.~E., 
Bianchi, L., Thilker, D.~A., et al.\ 2005, \apjl, 619, L87 

\bibitem[Hicken et al.(2009)]{hicken_constitution_2009} Hicken, M., Wood-Vasey,
W.~M., Blondin, S., et al.\ 2009, \apj, 700, 1097

\bibitem[Hicken et al.(2009)b]{hicken_H09_2009} Hicken, M., Challis, P.,
Jha, S., et al.\ 2009, \apj, 700, 331

\bibitem[Hinshaw et al.(2013)]{hinshaw_WMAP9} Hinshaw, G., Larson,
D., Komatsu, E., et al.\ 2013, \apjs, 208, 19

\bibitem[Hao et al.(2011)]{hao_convert_2011} Hao, C.-N., Kennicutt,
R.~C., Johnson, B.~D., et al.\ 2011, \apj, 741, 124

\bibitem[H{\"o}flich et al.(1998)]{hoeflich_1998} H{\"o}flich, P., 
Wheeler, J.~C., \& Thielemann, F.~K.\ 1998, \apj, 495, 617 

\bibitem[Humphreys et al.(2013)]{humphreys_correction_2013} Humphreys, E.~M.~L., 
Reid, M.~J., Moran, J.~M., Greenhill, L.~J., 
\& Argon, A.~L.\ 2013, \apj, 775, 13

\bibitem[Hunter et al.(2010)]{hunter_dwarfs_2010} Hunter, D.~A., 
Elmegreen, B.~G., \& Ludka, B.~C.\ 2010, \aj, 139, 447 

\bibitem[Jha et al.(2007)]{Jha_mlcs} Jha, S., Riess, A.~G., 
\& Kirshner, R.~P.\ 2007, \apj, 659, 122 


\bibitem[Kasen et al.(2009)]{kasen_2009} Kasen, D., R{\"o}pke, 
F.~K., \& Woosley, S.~E.\ 2009, \nat, 460, 869 

\bibitem[Kaviraj et al.(2012)]{kaviraj_tidal_2012} Kaviraj, S., Darg, D.,
Lintott, C., Schawinski, K., \& Silk, J.\ 2012, \mnras, 419, 70

\bibitem[{Kelly {et~al.}(2010)}]{kelly_hubble_2010}
Kelly, P.~L., Hicken, M., Burke, D.~L., Mandel, K.~S., \& Kirshner, R.~P. 2010,
\apj, 715, 743

\bibitem[Kennicutt et al.(2009)]{kennicutt_09} Kennicutt, R.~C.,
Jr., Hao, C.-N., Calzetti, D., et al.\ 2009, \apj, 703, 1672

\bibitem[Kessler et al.(2009)]{kessler_cosmo_2009} Kessler, R., Becker,
A.~C., Cinabro, D., et al.\ 2009, \apjs, 185, 32

\bibitem[Kewley et al.(2006)]{kewley_AGN_2006} Kewley, L.~J., Groves, 
B., Kauffmann, G., \& Heckman, T.\ 2006, \mnras, 372, 961 

\bibitem[Kim et al.(2014)]{akim14} Kim, A.~G., Aldering, G., 
Antilogus, P., et al.\ 2014, \apj, 784, 51 


\bibitem[{Lampeitl {et~al.}(2010)}]{lampeitl_effect_2010}
Lampeitl, H., Smith, M., Nichol, R.~C., {et~al.} 2010, ApJ, 722, 566

\bibitem[Lee et al.(2009)]{lee_SF_2009} Lee, J.~C., Gil de Paz, A.,
Tremonti, C., et al.\ 2009, \apj, 706, 599

\bibitem[Lee et al.(2011)]{lee_SF_2011} Lee, J.~C., Gil de Paz, A.,
Kennicutt, R.~C., Jr., et al.\ 2011, \apjs, 192, 6

\bibitem[Leitherer et al.(1999)]{leitherer_starburst99_1999} Leitherer, C., 
Schaerer, D., Goldader, J.~D., et al.\ 1999, \apjs, 123, 3 

\bibitem[Maoz et al.(2012)]{maoz_dtd} Maoz, D., Mannucci, F., 
\& Brandt, T.~D.\ 2012, \mnras, 426, 3282 


\bibitem[Marino et al.(2011)]{marino_Ecore_2011} Marino, A., Rampazzo, 
R., Bianchi, L., et al.\ 2011, \mnras, 411, 311 

\bibitem[Morrissey et al.(2007)]{GALEX_calibration} Morrissey, P.,
Conrow, T., Barlow, T.~A., et al.\ 2007, \apjs, 173, 682


\bibitem[Neff et al.(2005)]{neff_tidal_2005} Neff, S.~G., Thilker,
D.~A., Seibert, M., et al.\ 2005, \apjl, 619, L91

\bibitem[{Neill {et~al.}(2009)}]{neill_local_2009}
Neill, J.~D., Sullivan, M., Howell, D.~A., {et~al.} 2009, \apj, 707, 1449


\bibitem[O'Donnell(1994)]{odonnell_dust_law} O'Donnell, J.~E.\ 1994, 
\apj, 422, 158 

\bibitem[{Oey {et~al.}(2007)}]{oey_survey_2007}
Oey, M.~S., Meurer, G.~R., Yelda, S., {et~al.} 2007, \apj, 661, 801


\bibitem[Pan et al.(2014)]{pan_host_2014} Pan, Y.-C., Sullivan, M.,
Maguire, K., et al.\ 2014, \mnras, 438, 1391

\bibitem[{Perlmutter {et~al.}(1999)}]{perlmutter_measurements_1999}
Perlmutter, S., Aldering, G., Goldhaber, G., {et~al.} 1999, \apj, 517, 565

\bibitem[Planck Collaboration et 
al.(2014)]{planck_collaboration_planck_2013} Planck Collaboration, Ade, P.~A.~R.,
Aghanim, N., et al.\ 2014, \aap, 571, AA16 

\bibitem[Portegies Zwart et al.(2010)]{pz_2010} 
Portegies Zwart, S.~F., McMillan, S.~L.~W., \& Gieles, M.\ 2010, \araa, 48, 431 


\bibitem[Raskin et al.(2009)]{raskin_2009} Raskin, C., Scannapieco, 
E., Rhoads, J., \& Della Valle, M.\ 2009, \apj, 707, 74 

\bibitem[Rest et al.(2014)]{rest_cosmo_2013} Rest, A., Scolnic, D., 
Foley, R.~J., et al.\ 2014, \apj, 795, 44 

\bibitem[{Riess {et~al.}(1998)}]{riess_observational_1998}
Riess, A.~G., Filippenko, A.~V., Challis, P., {et~al.} 1998, \aj, 116, 1009

\bibitem[Riess et al.(2009)]{riess_h0_2009} Riess, A.~G., Macri, L., 
Casertano, S., et al.\ 2009, \apj, 699, 539 

\bibitem[{Riess {et~al.}(2011)}]{riess_3_2011}
Riess, A.~G., Macri, L., Casertano, S., {et~al.} 2011, \apj, 730, 119

\bibitem[{Rigault {et~al.}(2013)}]{rigault_evidence_2013}
Rigault, M., Copin, Y., Aldering, G., {et~al.} 2013, \aap, 560, A66

\bibitem[R{\"o}ser et al.(2010)]{roser_2010} R{\"o}ser, S., 
Kharchenko, N.~V., Piskunov, A.~E., et al.\ 2010, Astronomische 
Nachrichten, 331, 519

\bibitem[Roskar et al.(2008a)]{2008ApJ...675L..65R} Roskar, R.,
Debattista, V.~P., Stinson, G.~S., et al.\ 2008a, \apjl, 675, L65

\bibitem[Roskar et al.(2008b)]{2008ApJ...684L..79R} Roskar, R.,
Debattista, V.~P., Quinn, T.~R., Stinson, G.~S.,
\& Wadsley, J.\ 2008b, \apjl, 684, L79

\bibitem[Roskar et al.(2012)]{2012MNRAS.426.2089R} Ro{\v s}kar, R.,
Debattista, V.~P., Quinn, T.~R., \& Wadsley, J.\ 2012, \mnras, 426, 2089 


\bibitem[Salim et al.(2005)]{salim_prior_2005} Salim, S., Charlot, S., 
Rich, R.~M., et al.\ 2005, \apjl, 619, L39 

\bibitem[Salim et al.(2007)]{salim_uv_2007} Salim, S., Rich, R.~M.,
Charlot, S., et al.\ 2007, \apjs, 173, 267

\bibitem[Schawinski(2009)]{schawinsky_SFinETG_2009} Schawinski, K.\ 2009, 
\mnras, 397, 717 

\bibitem[Schlegel et al.(1998)]{schlegel_1998} Schlegel, D.~J., 
Finkbeiner, D.~P., \& Davis, M.\ 1998, \apj, 500, 525 

\bibitem[Seon et al.(2011)]{diffuse_FUV_Galaxy} Seon, K.-I., Witt, A., 
Kim, I.-J., et al.\ 2011, \apj, 743, 188

\bibitem[{{Seon} \& {Witt}(2012)}]{seon_witt_2012}
{Seon}, K.-I. \& {Witt}, A.~N. 2012, \apj, 758, 109

\bibitem[Scalzo et al.(2012)]{scalzo_91T} Scalzo, R., Aldering,
G., Antilogus, P., et al.\ 2012, \apj, 757, 12

\bibitem[Silverman et al.(2012)]{silverman_classification} Silverman, J.~M.,
Kong, J.~J., \& Filippenko, A.~V.\ 2012, \mnras, 425, 1819

\bibitem[Simones et al.(2014)]{simones_m31_2014} Simones, J.~E., Weisz, 
D.~R., Skillman, E.~D., et al.\ 2014, \apj, 788, 12 

\bibitem[Smith et al.(2010)]{smith_tidal_2010} Smith, B.~J., Giroux, 
M.~L., Struck, C., \& Hancock, M.\ 2010, \aj, 139, 1212 

\bibitem[Spergel et al.(2013)]{spergel_New_Planck} Spergel, D., Flauger,
R., \& Hlozek, R.\ 2013, arXiv:1312.3313

\bibitem[Strolger et al.(2002)]{strolger_99aw_2002} Strolger, L.-G., 
Smith, R.~C., Suntzeff, N.~B., et al.\ 2002, \aj, 124, 2905 

\bibitem[Sullivan et al.(2000)]{sullivan_SFRUV_2000} Sullivan, M., Treyer,
M.~A., Ellis, R.~S., et al.\ 2000, \mnras, 312, 442

\bibitem[Sullivan et al.(2010)]{sullivan_host_2010} Sullivan, M., Conley,
A., Howell, D.~A., et al.\ 2010, \mnras, 406, 782


\bibitem[Sullivan et al.(2011)]{sullivan_cosmo_2011} Sullivan, M., Guy, J., 
Conley, A., et al.\ 2011, \apj, 737, 102

\bibitem[Thilker et al.(2005)]{diffuse_FUV_M33} Thilker, D.~A., Hoopes, 
C.~G., Bianchi, L., et al.\ 2005, \apjl, 619, L67

\bibitem[Timmes et al.(2003)]{timmes_2003} Timmes, F.~X., Brown, 
E.~F., \& Truran, J.~W.\ 2003, \apjl, 590, L83 

\bibitem[Turner(1996)]{turner_cepheid_review} Turner, D.~G.\ 1996, \jrasc, 
90, 82 


\bibitem[van den Bergh et al.(2005)]{vdbergh_2005} van den Bergh, 
S., Li, W., \& Filippenko, A.~V.\ 2005, \pasp, 117, 773 

\bibitem[Veilleux \& Osterbrock(1987)]{veilleux_AGN_1987} 
Veilleux, S., \& Osterbrock, D.~E.\ 1987, \apjs, 63, 295 

\bibitem[Verley et al.(2010)]{verley_m33_2010}
Verley, S., Corbelli, E., Giovanardi, C., \& Hunt, L.~K.\ 2010, \aap, 510, A64

\bibitem[Whitmore et al.(1999)]{whitmore_cluster_ages_1999} Whitmore, B.~C., 
Zhang, Q., Leitherer, C., et al.\ 1999, \aj, 118, 1551 

\end{thebibliography}
